\documentclass[12pt]{article}
\usepackage{authblk}

\title{On Unitarity of the Hypergeometric Amplitude}
\author[1]{Gareth Mansfield}
\author[1,2]{Marcus Spradlin}
\date{}

\affil[1]{Department of Physics, Brown University, Providence, RI 02912, USA}
\affil[2]{Brown Theoretical Physics Center, Brown University, Providence, RI 02912, USA}

\usepackage[shortlabels]{enumitem}
\usepackage{amsmath}
\usepackage{amssymb}
\usepackage{amsthm}
\usepackage{graphicx}
\usepackage{tikz}
\usetikzlibrary{angles, quotes, hobby}
\usepackage{pgfplots}
\usepackage{physics}
\usepackage{fancyhdr}
\usepackage{pdflscape}
\usepackage{tcolorbox}
\usepackage{rotating}
\usepackage{cite}
\usepackage[linktocpage]{hyperref}

\def\R{\mathbb R}
\def\N{\mathbb N}
\def\Z{\mathbb Z}

\newcommand{\mc}{\mathcal}

\newtheorem{lemma}{Lemma}
\newtheorem{theorem}{Theorem}

\newtheorem{conjecture}[theorem]{Conjecture}

\pgfplotsset{compat=1.17}

\tcbset{
    colframe=green!50!blue,
    colback=green!50!blue!5,
    coltitle=white!20!,
    }

\numberwithin{equation}{section}

\hypersetup{
    colorlinks,
    linkcolor={blue},
    citecolor={blue},
    urlcolor={blue}
}

\begin{document}

\maketitle
\thispagestyle{empty} 
\setcounter{page}{0}

\begin{abstract}
    The hypergeometric amplitude is a one-parameter deformation of the Veneziano amplitude for four-point tachyon scattering in bosonic string theory that is consistent with $S$-matrix bootstrap constraints. In this article we construct a similar hypergeometric generalization of the Veneziano amplitude for type-I superstring theory. We then rule out a large region of the $(r,m^2,D)$ parameter space as non-unitary, and establish another large subset of the $(r, m^2, D)$ parameter space where all of the residue's partial wave coefficients are positive. We also analyze positivity in various limits and special cases. As a corollary to our analysis, we are able to directly demonstrate positivity of a wider set of Veneziano amplitude partial wave coefficients than what has been presented elsewhere.
\end{abstract}

\break

\tableofcontents

\section{Introduction}

In recent years there has been a resurgence of interest in the $S$-matrix bootstrap. This program aims to constrain properties of QFTs and string theories from first principles including Poincar\'e invariance, causality, locality, crossing symmetry, soft UV behavior, and unitarity \cite{Elvang:2015rqa,Cheung:2017pzi,Arkani-Hamed:2017jhn,Benincasa:2007xk,Cohen:2010mi}. In particular, following \cite{Caron-Huot:2016icg} many recent works have focused on meromorphic, i.e., tree-level, amplitudes. The Veneziano amplitude describing the tree-level $2\to 2$ scattering of open strings famously satisfies all of the abovementioned principles \cite{Veneziano:1968yb,Virasoro:1969me,Shapiro:1970gy}, but recent work has explored the extent to which certain variations thereof may also do so. It is interesting to wonder whether string theory can be derived from a set of basic physical principles, or whether more amplitudes exist that satisfy the same constraints? Many 4-scalar amplitudes have been constructed and rigorously demonstrated to satisfy many of the bootstrap constraints \cite{Coon:1969yw, Cheung:2022mkw,Cheung:2023adk, Cheung:2023uwn, Haring:2023zwu, Bhardwaj:2023eus, Huang:2022mdb,Geiser:2022exp,Geiser:2022icl, Geiser:2023qqq, Eckner:2024ggx,Cheung:2024uhn, Cheung:2024obl}, and much work has been done aiming to bootstrap string theory directly \cite{Caron-Huot:2016icg,Cheung:2024uhn,Cheung:2024obl,Berman:2023jys,Berman:2024wyt,Chiang:2023quf,Arkani-Hamed:2023jwn,Albert:2024yap}. For a more comprehensive overview of recent literature in this area, see the introduction of \cite{Haring:2023zwu} and references therein. 

\hspace{0pt}

\begin{figure}[ht]
    \centering
    \includegraphics[width=8.1cm]{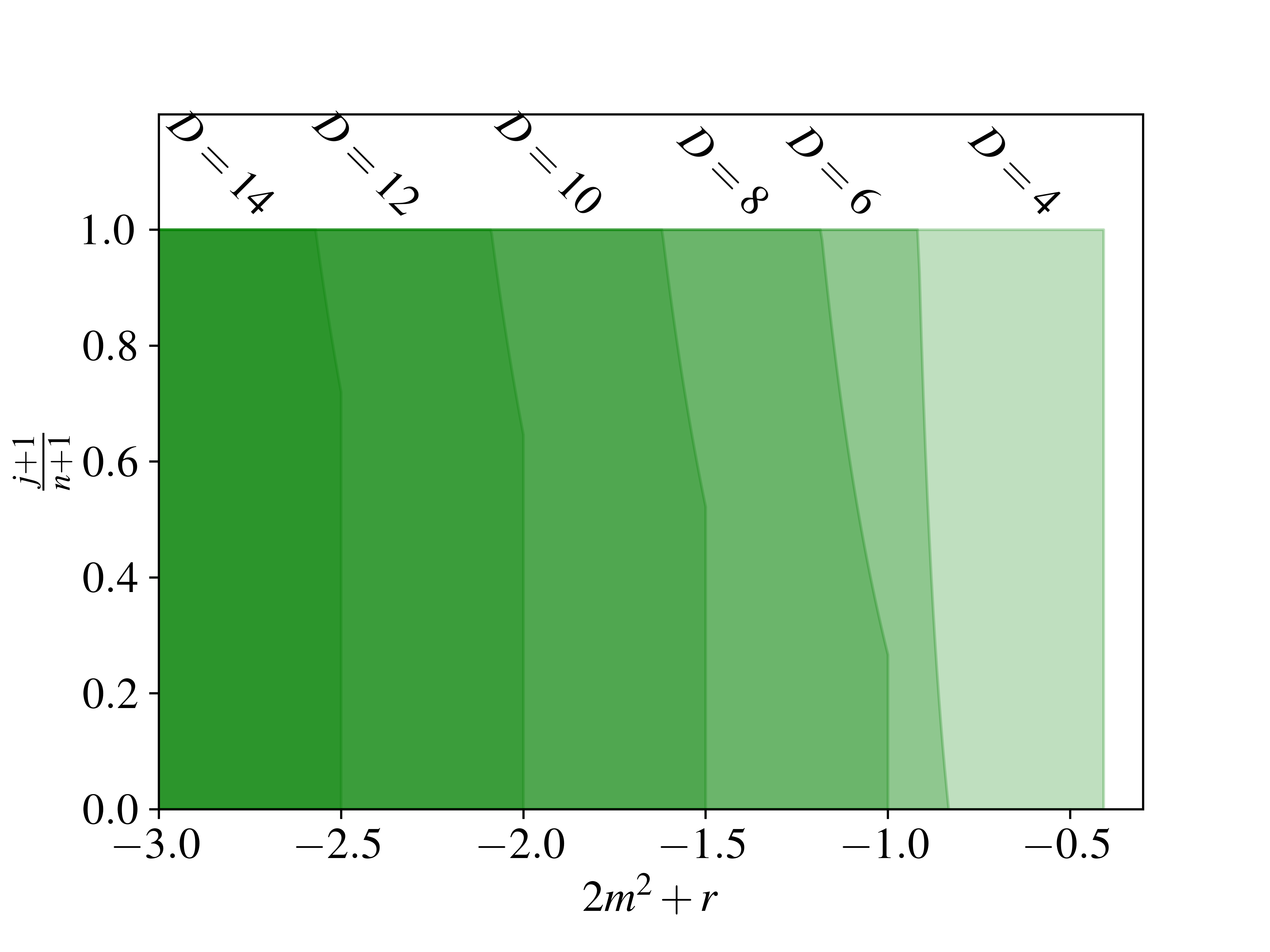}
    \includegraphics[width=8.1cm]{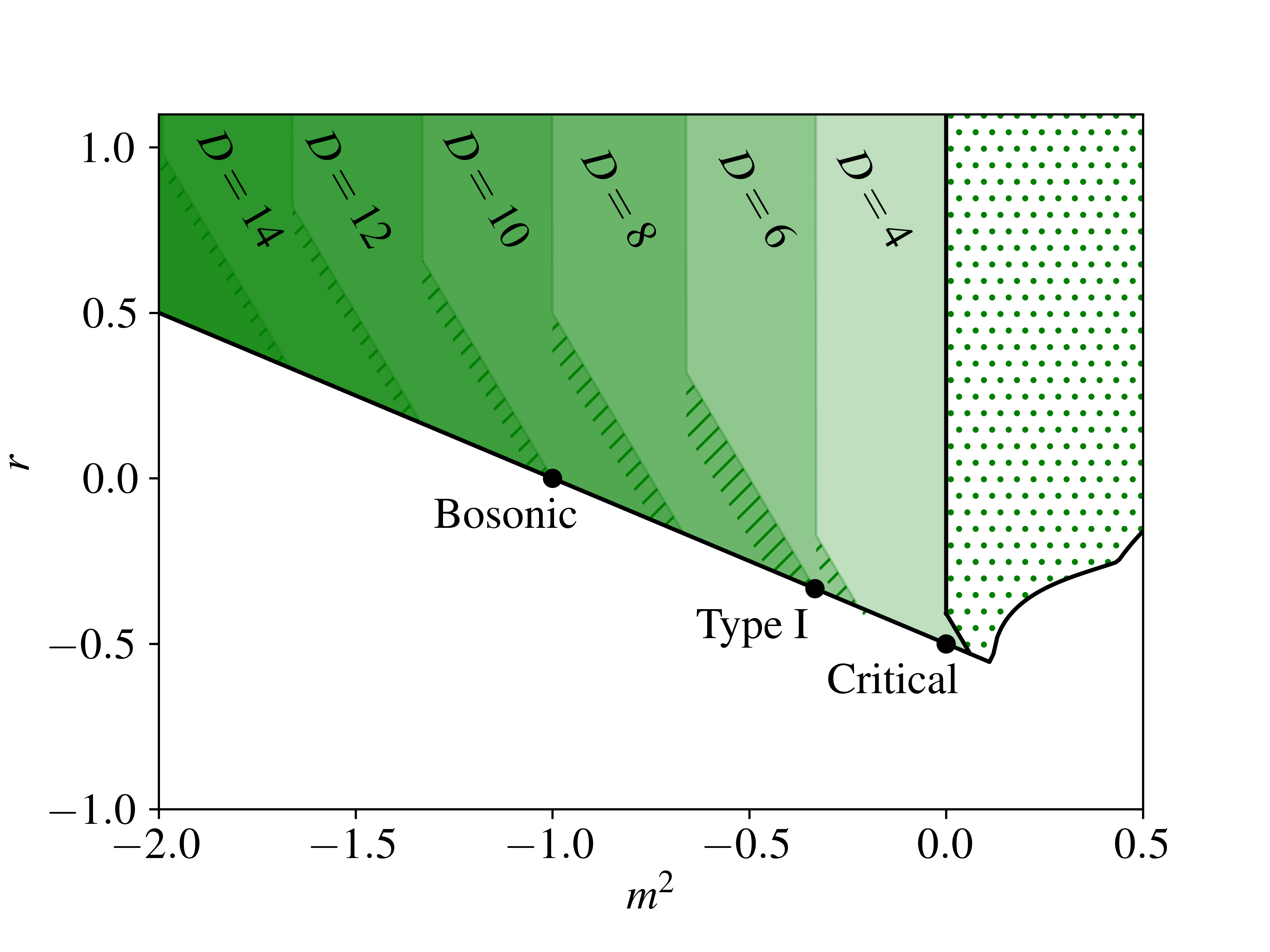}
    \caption{Positivity bounds obtained by combining theorems \ref{theorem3}, \ref{theorem4}, and \ref{theorem5} at different $D$. The left figure plots $2m^2+r$ against $(j+1)/(n+1)$. We see that when $D\neq 4$, the top right corner of each positive region is ``cut out'', indicating the slim region where lemma 2 is too weak to guarantee positivity. The right figure plots the same positivity regions in $m^2-r$ space. The solid green regions represent areas where $B_{n,j}\geq 0$ for all $n,j$, and the ``cut out" areas where lemma 2 fails are represented by the striped regions.}
    \label{orangepurple}
\end{figure}

At tree level, unitarity of the $S$-matrix implies the constraint that the residues of the amplitude admit a decomposition into a positive sum of spherical harmonics, and this is difficult to verify analytically. The amplitudes corresponding to string theory are only indirectly 
understood to satisfy this positivity constraint via the no-ghost theorem \cite{Goddard:1972iy}, and for more general amplitudes, most cases in the literature resort to numerical tests of positivity. There is significant interest in a more direct and rigorous treatment of positivity \cite{Green:2019tpt,Maity:2021obe,Arkani-Hamed:2022gsa, Chakravarty:2022vrp, Figueroa:2022onw,Bhardwaj:2022lbz,Jepsen:2023sia, Bhardwaj:2024klc, Rigatos:2023asb,Eckner:2024ggx,Rigatos:2024beq,Wang:2024wcc}. In particular, recent work has been done on the positivity of the hypergeometric amplitude introduced in \cite{Cheung:2023adk}. The original paper \cite{Cheung:2023adk} provides a detailed numerical analysis which carves out a region in parameter space that is expected to be unitary. In \cite{Rigatos:2023asb}, the authors developed an explicit formula for the residue's partial wave coefficients at $m^2=-1$ that ruled out certain non-unitary regions of parameter space analytically. The authors of \cite{Rigatos:2024beq,Wang:2024wcc} developed a representation of the $q$-hypergeometric partial-wave coefficients in a basis of ``cyclotomic $q$-deformed harmonic numbers'', which at $q=1$ demonstrated the positivity of a set of low-spin coefficients for the hypergeometric amplitude. In this article, we restrict to the $q=1$ case of the hypergeometric amplitude, and provide a complete analytical proof of positivity for all of the residues' partial wave coefficients in a large subset of the $(r,m^2,D)$ parameter space\footnote{The bulk of this region is established in theorem \ref{theorem4} and is depicted in Fig. \ref{fig:green}. Theorems \ref{theorem5}, \ref{theorem6}, and \ref{theorem7} make the region slightly larger, as depicted in Fig. \ref{orangepurple}.} by generalizing a double-contour representation of the Veneziano amplitude developed in \cite{Arkani-Hamed:2022gsa} out to nonzero $r$. We additionally use this formula to demonstrate positivity at low spin and at large mass levels. In doing so, we support that the hypergeometric amplitude is a well-behaved alternative to the Veneziano amplitude for four-particle scattering. We emphasize that positivity is a necessary-but-insufficient condition to enforce unitarity; unitarity of the full amplitude demands both positivity for four-particle scattering and a generalization to $N>4$ particles, and currently no such higher-point generalization exists. Therefore areas in parameter space where we demonstrate violations of positivity mean that the amplitude is non-unitary, whereas areas where positivity is satisfied could also still later be demonstrated to be non-unitary if they are found not to admit a consistent higher-point generalization.

\hspace{0pt}

In Section \ref{sec:background}, we review the definition of the hypergeometric amplitude, construct a new superstring generalization, and explain some important properties. In Section \ref{sec:direct}, we provide an overview of several positivity bounds that can be seen directly from the explicit expression of the residue's partial wave coefficients as an inner product with a Gegenbauer polynomial. In Section \ref{sec:contour}, we introduce a contour integral formula for the residue's partial wave coefficients, and use it to carve out a region in $(r,m^2,D)$ parameter space where all coefficients are positive. We also find larger positive regions when specific restrictions are placed on $n$ and $j$. In Section \ref{sec:limiting}, we develop asymptotic formulas for the coefficients that demonstrate positivity along fixed-spin and Regge trajectories at large $n$, and we also reproduce a result of \cite{Rigatos:2024beq,Wang:2024wcc} that demonstrates positivity in some special low-spin cases. We conclude in Section \ref{sec:conclusion} with a condensed summary of all presented positivity bounds and a detailed analysis of how they overlap with previous results.

\section{Background}
\label{sec:background}

The hypergeometric amplitude constructed in \cite{Cheung:2023adk} is a crossing-symmetric, UV-complete amplitude that depends on two real parameters: $m^2$ and $r$. It takes the form
\begin{align}
    \mc A^1(s,t)=-\frac1{s-m^2} {}_3F_2\left[\begin{matrix}1,\>-s+m^2,\>1+t+r-m^2\\1- s+m^2,\>1+r\end{matrix};1\right]\label{eq:hg}
\end{align}
where ${}_3 F_2$ is the generalized hypergeometric function \cite{alma991033988579703276}. Sending $r\to 0$ reduces the amplitude to
\begin{align}
    \mc A^1(s,t)\stackrel{r\to 0}{=}\frac{\Gamma(m^2-s)\Gamma(m^2-t)}{\Gamma(2m^2-s-t)},\label{veneziano}
\end{align}
which coincides with the Veneziano amplitude for bosonic string theory when $m^2=-1$. The ``1" superscript denotes that $\mc A^1$ is a deformation of the bosonic Veneziano amplitude, distinguishing it from the superstring version that will be introduced in (\ref{eq:hgnew}). $\mc A^1$ can be written in a manifestly crossing-symmetric form by performing a Thomae transformation \cite{alma991033988579703276} on (\ref{eq:hg}) to obtain
\begin{align}
    \mc A^1(s,t)=\frac{\Gamma(m^2-s)\Gamma(m^2-t)}{\Gamma(2m^2-s-t)}{}_3F_2\left[\begin{matrix}-s+m^2,\>-t+m^2,\>r\\-s-t+2m^2,\>1+r\end{matrix};1\right].
\end{align}

$\mc A^1$ is meromorphic with simple poles at each $s=m^2+n$, $n=0,1,2,\ldots$ which gives $m^2$ an interpretation as the mass gap of the theory. (\ref{eq:hg}) is also consistent with locality, as the residue at each pole is a polynomial in $t$; we define
\begin{align}
    R_n^1(t)=\Res_{s=m^2+n} \mc A^1(s,t)=\frac{(1+t+r-m^2)_{(n)}}{(r+1)_{(n)}}\label{eq:res}
\end{align}
where $a_{(n)}=a(a+1)\ldots (a+n-1)$ denotes the rising factorial. 

\hspace{0pt}

Unitarity of the $S$-matrix requires that the residue polynomials may be decomposed into the partial-wave expansion
\begin{align}
    R_n(t)=\sum_{j=0}^nB_{n,j} G^D_j(\cos\theta)
\end{align}
where $\theta$ is the scattering angle for particles of mass $m^2$, $G_j^D$ is the $j$th $D$-dimensional Gegenbauer polynomial, and $B_{n,j}\geq 0$ for all $j,n$ where $n\geq 3m^2$.\footnote{The $n\geq 3m^2$ constraint arises from a requirement that the energy $m^2+n$ of the exchanged particle must exceed the total mass $4m^2$ of the external states, as emphasized in \cite{Arkani-Hamed:2023jwn,Cheung:2024uhn}.} We say that an amplitude satisfies positivity when this condition holds. The authors of \cite{Cheung:2023adk,Rigatos:2023asb,Rigatos:2024beq,  Wang:2024wcc} have done previous work exploring this positivity condition in a large region of parameter space.

\hspace{0pt}

As pointed out in \cite{Cheung:2023adk}, one immediate avenue for demonstrating positivity comes from the following worldsheet integral representation of the hypergeometric amplitude:
\begin{align}
    \mc A^1(s,t)=r\int_0^1\int_0^1\dd x\>\dd y\> \frac{x^{-s+m^2-1}y^{r-1}(1-xy)^{t-m^2}}{(1-x)^{t-m^2+1}}.\label{eq:intformula}
\end{align}
This is equivalent to the 5-point Veneziano amplitude for open bosonic strings, with Mandelstam variables fixed to $s_{12}=s-m^2$, $s_{23}=t-m^2$, $s_{34}=s_{51}=-1$, and $s_{45}=-r$. This amplitude is known to be unitary at $m^2=-1$ and $D\leq 26$ via the no-ghost theorem; therefore, the hypergeometric amplitude satisfies positivity at $m^2=-1$, $D\leq 26$.

\hspace{0pt}

Now we introduce a new amplitude that generalizes the 4-point type-I superstring amplitude in a manner similar to the way (\ref{eq:hg}) generalizes the bosonic string amplitude. Using shifted Mandelstam invariants $\bar s=s-m^2$, $\bar t=t-m^2$ for convenience, we define 
\begin{align}
    \mc A^0(s,t)=\frac{1}{\bar s \bar t}-\frac{1}{(1+r)(1-\bar s)}\>{}_3 F_2\left[\begin{matrix}1,\>1- \bar s,\> 1+\bar t+r\\2-\bar s,\>2+r\end{matrix};1\right].\label{eq:hgnew}
\end{align}
This function can be obtained from the original hypergeometric amplitude by shifting the inputs and adding an extra pole. 
We note that the factor of $1/(r+1)$ could be replaced by any function $f(r)$ that satisfies $f(0)=1$; we have chosen $f(r)=1/(r+1)$ so that later on in equation (\ref{eq:newres}), the denominator matches that of (\ref{eq:res}). $\mc A^0(s,t)$ admits the following dual-resonant representation:
\begin{align}
    \mc A^0(s,t)=\frac1{\bar s\bar t}+\sum_{n=1}^{\infty}\frac{1}{n-\bar s}\frac{(1+\bar t+r)_{(n-1)}}{(r+1)_{(n)}}.\label{eq:newhg}
\end{align}
From this form we see that $\mc A^0(s,t)$ is aesthetically similar to the amplitude $\mc A^1(s,t)$ discussed above: it has simple poles at each $s=m^2+n$ with $n=0,1,2,\ldots$ and the residues for $n\geq 1$ are given by the polynomials
\begin{align}
    R_n^0(t)=\Res_{s=m^2+n}\mc A^0(s,t)=\frac{(1+t+r-m^2)_{(n-1)}}{(r+1)_{(n)}},\label{eq:newres}
\end{align}
which differ from the residues in (\ref{eq:res}) only by a factor of $\bar t+r+n$. Crossing symmetry in $s\leftrightarrow t$ can be made manifest by performing a Thomae transformation on (\ref{eq:hgnew}) to obtain
\begin{align}
     \mc A^0(s,t)=\frac 1{\bar s\bar t}-\frac{1}{r+1}\frac{\Gamma(1-\bar s)\Gamma(1-\bar t)}{\Gamma(2-\bar s-\bar t)}\>{}_3 F_2\left[\begin{matrix}1-\bar t,\>1- \bar s,\> 1+r\\2-\bar s-\bar t,\>2+r\end{matrix};1\right].
\end{align}
And lastly, we can compute the $r\to 0$ limit:
\begin{align}
    \mc A^0(s,t)\stackrel{r\to 0}{=}\frac{\Gamma(m^2-s)\Gamma(m^2-t)}{\Gamma(1+2m^2-s-t)}.\label{typeI}
\end{align}
When $m^2=0$, this coincides with the Veneziano amplitude for scattering in type-I string theory. So (\ref{eq:hgnew}) is a deformation of the type-I Veneziano amplitude, just as (\ref{eq:hg}) is a deformation of the bosonic Veneziano amplitude. Throughout the paper we will use the notation $\mc A^a$, $R^a_n(t)$, and $B_{n,j}^a$ with $a\in\{0,1\}$ to distinguish results for the bosonic ($a=1$) and type-I ($a=0$) amplitudes.

\hspace{0pt}

\begin{tcolorbox}[title= \textbf{Note about conventions}]

    In many sources, the Veneziano amplitude is defined as
    \begin{align}
        \frac{\Gamma(-s)\Gamma(-t)}{\Gamma(-s-t)},
    \end{align}
    which is also known as the four-point amplitude of $Z$-theory~\cite{Carrasco:2016ldy} and corresponds to the $r=0,\>a=1,\>m^2=0$ case of our hypergeometric amplitude. However, this amplitude does not  describe the scattering of bosonic nor type-I strings. Other authors define it in terms of the shifted variables 
    \begin{align}
        \frac{\Gamma(- s+m^2)\Gamma(- t+m^2)}{\Gamma(2m^2- s- t)},
    \end{align}
    which corresponds to the $r=0,\> a=1$ case of our amplitude. This correctly produces the bosonic Veneziano amplitude at $m^2=-1$, but does not produce the type-I amplitude at $m^2=0$. These conventions can be misleading and we do not refer to either of them when we discuss the Veneziano amplitude; we instead refer to equations (\ref{veneziano}) at $m^2=-1$ and (\ref{typeI}) at $m^2=0$ for bosonic and type-I strings respectively. However, these alternate definitions do lie within the parameter space of our model, so many of our results can still be applied to these functions with minimal effort.
\end{tcolorbox}

\section{Direct Positivity Constraints}
\label{sec:direct}
To extract the $j$th Gegenbauer coefficient $B^a_{n,j}$, we use the Rodrigues formula for $G_j^D(x)$ and integrate by parts:
\begin{align}
    B^a_{n,j}&= \int_{-1}^1\dd x\> (1-x^2)^\delta G^D_j(x)R^a_n(t)= \alpha_{j,D}\int_{-1}^1\dd x\>(1-x^2)^{\delta+j}\dv[j]{x}R^a_n(t),
\end{align}
where $\delta=(D-4)/2$, $\alpha_{j,D}$ is a positive normalization constant, and $x$ and $t$ are related by
\begin{align}
    t=\frac{(s-4m^2)(x-1)}{2}=\frac{(n-3m^2)(x-1)}{2}.
\end{align}
This gives the complete result
\begin{align}
B^a_{n,j}&=\frac{\alpha_{j,D}}{(r+1)_{(n)}}\int_{-1}^{1}\dd x\> (1-x^2)^{\delta+j}\pdv[j]{x} \left[\frac12(n-3m^2)x+\frac{m^2-n}{2}+r+1\right]_{(n-1+a)}.\label{eq:gegen}
\end{align}
In this section, we will present some preliminary results that can be obtained by direct examination of (\ref{eq:gegen}), before turning to the contour-integral representation in section \ref{sec:contour}. First, from (\ref{eq:gegen}), one can see the following relation between the bosonic ($a=1$) and type-I ($a=0$) coefficients:
\begin{align}
    B^0_{n,j}(r,m^2)=\frac{(r+\frac23)_{(n-1)}}{(r+1)_{(n)}}B^1_{n-1,j}\left(r-\frac13,m^2-\frac13\right).\label{eq:inv}
\end{align}
This relation drastically simplifies the positivity analysis by allowing us to easily transpose results between the two variants of the amplitude. It also tells us that positivity of the $a=0$ hypergeometric amplitude at $r=m^2=-\frac13$ is equivalent to positivity of the type-I Veneziano amplitude. Since this amplitude is known to be unitary in $D=10$ via the no-ghost theorem, we obtain positivity of the $a=1$ hypergeometric amplitude when $D\leq 10$, $r=m^2=-\frac13$. It would be interesting to see if this result can be extended to $r>-\frac13$ by developing an integral formula for the Type-I ($a=0$) hypergeometric amplitude akin to (\ref{eq:intformula}); we leave this for future work.

\subsection{Relations to Other Amplitudes}

It is worth mentioning another generalization of the Veneziano amplitude, given by
\begin{align}
    \mc B(s,t)=\frac{\Gamma(m^2-s)\Gamma(m^2-t)}{\Gamma(1-a+2m^2-s-t)}
\end{align}
for $a\in\mathbb N$, $m^2\in \R$. This ``satellite" amplitude is studied in \cite{Gross:1969db, Cheung:2022mkw}. Like the hypergeometric amplitude, it has simple poles at each $m^2+n$ for $n=0,1,\ldots$. The residues are given by the expression
\begin{align}
    \Res_{s=m^2+n}\mc B(s,t)=\frac{(1+t-m^2)_{(n+a-1)}}{n!},
\end{align}
which bears a remarkable resemblance to the hypergeometric residue in (\ref{eq:res}). This reflects a relation between the hypergeometric and satellite amplitudes that is described in more detail in \cite{Cheung:2023adk}. The partial wave coefficients of this amplitude's residues can be computed from those of the hypergeometric via
\begin{align}
    B_{n,j}^{\rm satellite}(a,m^2)=\frac{(r+1)_{(n-1+a)}}{n!}B_{n+a-1,j}^{1}\left(\frac {a-1}3,m^2+\frac {a-1}3\right).
\end{align}
Using this relation, one can extend all positivity results for the hypergeometric to the satellite amplitude as well.

\hspace{0pt}

Another generalization proposed by \cite{Cheung:2024obl} is the 1-parameter family of ``planar analogue" amplitudes
\begin{align}
\mc C(s, t)=-\frac{1}{s t}+\frac{\lambda \Gamma(\lambda-\lambda s) \Gamma(\lambda-\lambda t)}{\Gamma(2 \lambda-\lambda s-\lambda t)}{ }_3 F_2\left[\begin{array}{c}
\lambda(1-s), \lambda(1-t), \frac{3 \lambda-1}{2} \\
\lambda(2-s-t), \frac{3 \lambda+1}{2}
\end{array} ; 1\right]
\end{align}
for $\lambda\in[0,\infty)$. These are equivalent to the type-I hypergeometric amplitude with rescaled inputs. The partial wave coefficients of this amplitude's residues are related, up to a positive constant, to those of the type-I hypergeometric via
\begin{align}
    B^{\text{planar}}_{n,j}(\lambda)\propto B^0_{n,j}\left(\frac{\lambda-1}{6},\frac{1-\lambda}{3}\right),\label{eq:grav}
\end{align}
which corresponds to the diagonal line along the bottom of Fig. \ref{red}. Using (\ref{eq:grav}), one can extend many of the results in this paper to the gravitational amplitudes of \cite{Cheung:2024obl}.

\subsection{Non-Unitary Regions}
\label{sec:nonunitary}

The generalized hypergeometric amplitude violates positivity in a large subset of its parameter space. Here we provide a complete analytic derivation of this positivity bound, reproducing the numerical results and special cases studied in \cite{Cheung:2023adk, Rigatos:2023asb, Wang:2024wcc}. We split the bound into two regions, as the proof in each region is significantly different.

\begin{theorem}\label{theorem1}
    The generalized hypergeometric amplitude violates positivity in all $D\geq 4$ if $2r+m^2+a<0$, or if $r<-1$ and $\lfloor r\rfloor$ is even.
\end{theorem}
\textit{Proof.} 
If we fix a value of $n-j$, it is possible to explicitly compute (\ref{eq:gegen}) for the coefficients on leading trajectories. We examine the leading and sub-leading trajectories, $B^a_{n,n-1+a}$, $B^a_{n,n-2+a}$, obtaining (up to a positive factor) the following expressions:
\begin{align*}
    B^a_{n,n-1+a}\propto\frac{1}{(r+1)_{(n)}}&&B^a_{n,n-2+a}\propto \frac{2r+m^2+a}{(r+1)_{(n)}}
\end{align*}
Then we split into three cases and find a negative coefficient in each:
\begin{itemize}
    \item If $r<-1$ and $\lfloor r\rfloor$ is even, then $B^a_{n,n-1+a}< 0$ for any even $n> -r-1$. 
    \item If $r<-1$, $2r+m^2+a< 0$, and $\lfloor r\rfloor$ is odd, then $B^a_{n,n-2+a}<0$.
    \item If $r>-1$ and $2r+m^2+a< 0$, then $B^a_{n,n-2+a}<0$.
\end{itemize}
This completes the proof; note that in all cases we must choose $n\geq 3m^2$. Explicit calculations of more Regge trajectories, which provide stronger positivity bounds than those listed above, are present in Appendix \ref{sec:regge}. Beyond analysis of leading Regge trajectories, there is another way that we can work directly from  (\ref{eq:gegen}) to rule out much of the remaining parameter space:

\begin{theorem}\label{theorem2}
    The generalized hypergeometric amplitude violates positivity in all $D\geq 4$ if $m^2>r+1$ and $\lceil 6m^2\rceil$ is odd.
\end{theorem}

\textit{Proof.} For simplicity we assume $a=1$; the proof is nearly identical for the $a=0$ case. Change variables in (\ref{eq:ygegen}) to $y=\frac12(n-3m^2)x+\frac{m^2-n}{2}+r+1$ and write the residue's partial wave coefficients as
\begin{align}
B^1_{n,j}&\propto\frac1{(r+1)_{(n)}}\int_{-n+2m^2+r+1}^{-m^2+r+1}\dd y\> (1-x^2)^{\delta+j}\pdv[j]{y} y_{(n)}.\label{eq:ygegen}
\end{align}
If $(r+1)_{(n)}$ is negative, we immediately obtain negativity of $B^1_{n,n}$ for any even $n>-r-1$; therefore we will assume the prefactor $1/(r+1)_{(n)}$ is positive. Then we choose $n=\lceil 3m^2\rceil$, which is the smallest value of $n$ consistent with the $n\geq 3m^2$ requirement. Expand out the derivative of the rising factorial as
\begin{align}
    \pdv[j]{y} y_{(n)}=p^{(j)}(y)=c^j(y-\rho^j_0)\ldots(y-\rho^j_{n-j-1})\label{eq:roots}
\end{align}
for roots $0\geq \rho^j_0>\ldots>\rho^j_{n-j-1}$ and leading coefficient $c^j>0$. (\ref{eq:ygegen}) will be negative if the integration domain is a subset of $[\rho^j_1,\rho^j_0]$, since (\ref{eq:roots}) is negative throughout this interval. This motivates the choice of $j$ to be the largest number below $n$ for which $1-m^2+r\leq \rho^j_0$ (note that the $m^2>r+1$ requirement guarantees that $1-m^2+r< 0=\rho_0^0$). From here it suffices to show that the lower integration bound in (\ref{eq:ygegen}), which is bounded from below by $-n+2m^2+r+1\geq \frac12 -m^2+r$ due to the constraint that $\lceil 6m^2\rceil$ is odd, is greater than $\rho^j_1$. This will prove that the integration domain in (\ref{eq:ygegen}) is contained within $[\rho^j_1,\rho^j_0]$. To do this, we will need two important relations between the roots of $p^{(j)}(x)$:
\begin{enumerate}[(1)]
 \item $\rho^j_{k}-\rho^j_{k+1}\geq 1$. This is clearly true for $j=0$. Then proceeding by induction on $j$, it suffices to prove that there exists an $\ell$ such that $\rho^j_{k}-\rho^j_{k+1}\geq \rho^{j-1}_{\ell}-\rho^{j-1}_{\ell+1}$ for all $k$. We can expand out the derivatives of $p(x)$ in the form 
    \begin{align}
        p^{(j)}(x)=\frac{c^j}{c^{j-1}}p^{(j-1)}(x)\sum_{\ell=0}^{n-j-2} \frac{1}{\rho^{j-1}_\ell-x}.
    \end{align}
    Then using $0=p^{(j)}(\rho_k^j)-p^{(j)}(\rho_{k+1}^j)$, we can compute
    \begin{align}
        0=\frac{1}{\rho^{j-1}_{n-j-2}-\rho_k^j}-\frac{1}{\rho^{j-1}_{0}-\rho_{k+1}^j}+\sum_{\ell=0}^{n-j-3} \frac{1}{\rho^{j-1}_\ell-\rho_k^j}-\frac{1}{\rho^{j-1}_{\ell+1}-\rho_{k+1}^j}.
    \end{align}
    The first two terms are positive since all roots of $p^{(j)}(x)$ must lie on the interval $[\rho_{n-j-1}^{j-1},\rho_{0}^{j-1}]$. Therefore the sum over $\ell$ must contain at least one negative term to cancel these out. But this can only be true if for some $\ell$, $\rho_{k}^j-\rho_{k+1}^j\geq\rho^{j-1}_\ell-\rho^{j-1}_{\ell+1}$, which is what we wanted.
    
    \item $\rho_0^{j+1}\geq \frac{\rho_0^j+\rho_1^j}2$. Setting $p^{(j+1)}(\rho^{j+1}_0)=0$ and using the sum formula from (1), we have the bound
    \begin{align}
        \frac1{\rho^j_0-\rho^{j+1}_0}&=\sum_{k=1}^{n-j}\frac{1}{\rho^{j+1}_0-\rho^j_k}\geq \frac{1}{\rho^{j+1}_0-\rho^j_1},
    \end{align}
    which is equivalent to the desired result.
\end{enumerate}
Now for the sake of contradiction, assume $\frac12-m^2+r< \rho^j_1$. Then (1) and (2) demand that
\begin{align}
    \rho^{j+1}_0\geq \frac{\rho_0^j+\rho_1^j}{2}\geq\frac{2\rho_1^j+1}{2}\geq\frac{2(\frac12-m^2+r)+1}{2}=1-m^2+r,
\end{align}
but this contradicts our assumption that $j$ is the largest integer for which $\rho^j_0\geq 1-m^2+r$, proving the theorem. 

\hspace{0pt}

\begin{figure}[ht]
    \centering
    \includegraphics[width=12cm]{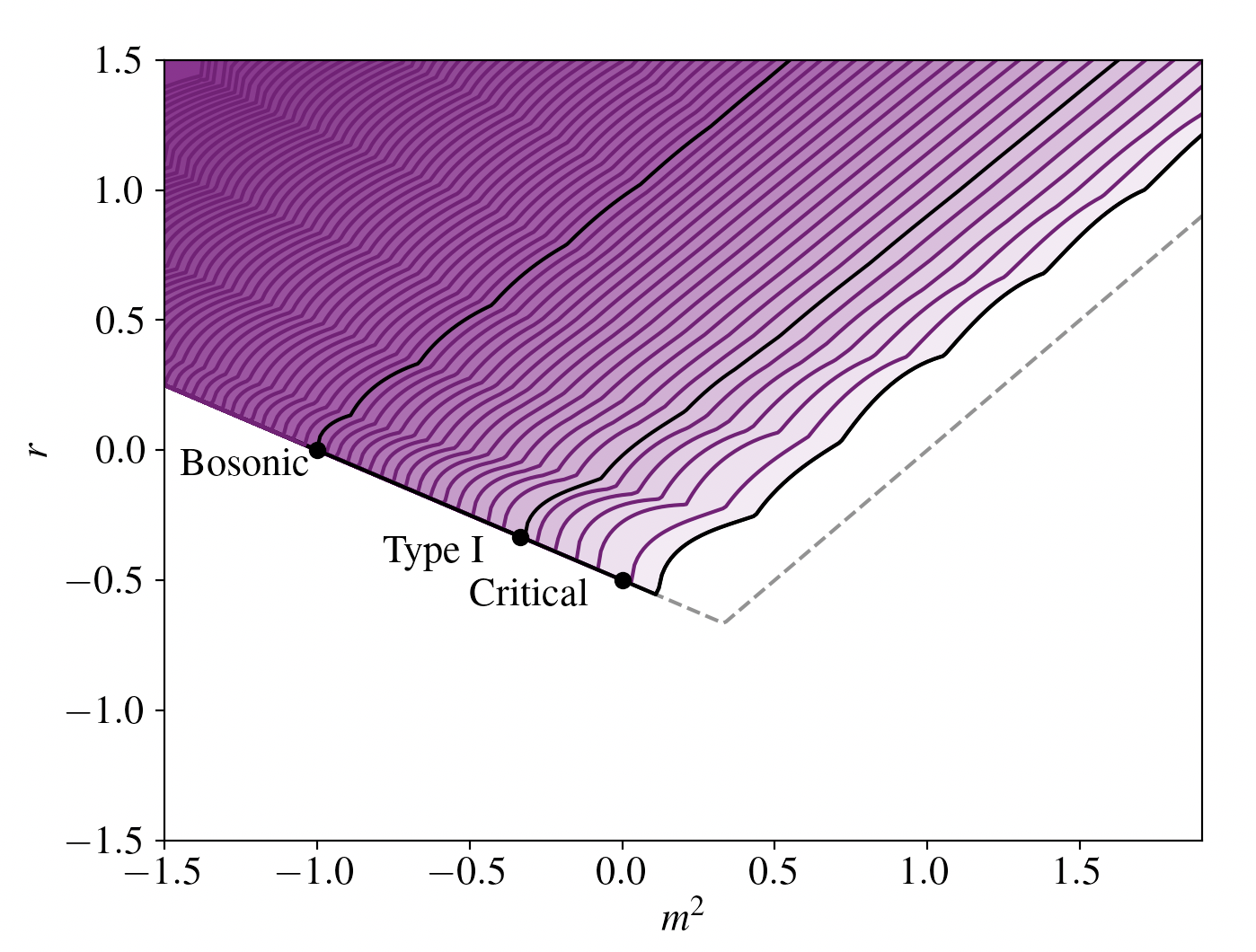}
 \caption{Positivity bounds from all coefficients $B_{n,j}^1$ at $3m^2\leq n\leq 10$ for different values of $D$. The white region is non-positive in all $D\geq 4$. The purple regions indicate where the studied coefficients only become negative when $D$ lies above some threshold; the three solid black lines indicate this threshold for $D=4$, 10, and 26. The dashed gray line indicates the analytical bound given by theorems \ref{theorem1}, \ref{theorem2}, and conjecture \ref{conjec}. Also noted are points in the parameter space corresponding to the bosonic Veneziano amplitude, an amplitude whose residues' partial wave coefficients differ by positive factors from those of the type-I Veneziano amplitude (in accordance with (\ref{eq:inv})), and a ``critical'' amplitude at $m^2=0$ whose critical dimension is exactly $D=5$. }
    \label{red}
\end{figure}

Theorems \ref{theorem1} and \ref{theorem2} immediately exclude a large subset of the parameter space, which is depicted in Fig. \ref{fig:negative} alongside more results that will be proved later. We can run numerical tests to exclude wider swaths of the parameter space as $D$ increases, which are depicted in Fig. \ref{red}.

\hspace{0pt}

We end this section with a conjecture that would eliminate the need for the constraint that $\lceil 6m^2\rceil$ is odd in theorem \ref{theorem2}, which would instead simply rule out all amplitudes at $m^2>r+1$. The $\lceil 6m^2\rceil$ constraint was necessary to ensure that the width of the integration domain in (\ref{eq:ygegen}), $[-\lceil 3m^2\rceil+2m^2+r+1,-m^2+r+1]$, was less than $\frac12$, making the interval small enough to guarantee that it is entirely contained by $[\rho_1,\rho_0]$. We can eliminate this requirement if the following theorem holds:

\hspace{0pt}

\begin{conjecture}\label{conjec}
    \textit{The roots $\rho_k^j$ as defined in (\ref{eq:roots}) satisfy $\rho_1^j-\rho_2^j\leq \rho_0^j-\rho_1^j$ for all $j$ and $n$, provided $n-j\geq 3$.}
\end{conjecture}

\hspace{0pt}

If this is the case, which numerical evidence suggests, then it would suffice for only the right half of the integration domain to fall within $[\rho_1,\rho_0]$, as one can use the conjecture to show that the right half of the integral of $p^{(j)}(y)$ will have a larger absolute value than the left half. This would guarantee negativity of the entire integral, and remove the restriction on $\lceil 6m^2\rceil$.

\subsection{\texorpdfstring{Positivity at large $r$}{Positivity at large r}}

There is a direct analytical way that we guarantee positivity of the residue's partial wave coefficients at sufficiently low mass level and/or sufficiently high spin by directly analyzing equation (\ref{eq:gegen}):

\begin{theorem}\label{theorem3}
    $B^1_{n,j}\geq 0$ if $r>-1$ and $n \leq 2+2r+m^2+ 2j(2H_{2n-3}-H_{n-1})^{-1}$ where $H_k$ denotes the $k$th harmonic number.
\end{theorem}

\textit{Proof.} The integrand in (\ref{eq:gegen}) can be split into two factors. The first factor, $(1-x^2)^{\delta+j}$, is symmetric about $x=0$. The second factor is a degree $n-j$ polynomial $p(x)$ with positive leading coefficient and real roots. The integral from $-1$ to $1$ of such a polynomial is guaranteed to be positive when all roots are non-positive. So to prove positivity, it suffices to show that the rightmost root of the polynomial in (\ref{eq:gegen}) is less than or equal to zero. When $j=0$, this root lies at exactly $\frac12(n-m^2)-r-1$, which immediately yields the bound stated in the theorem. For $j\geq 1$, we have the bound $\rho_0^j-\rho_0^{j+1}\geq (2H_{2n-3}-H_{n-1})^{-1}$ on the rightmost root, which follows from the inequality 
\begin{align}
    -\frac1{\rho^{j+1}_0-\rho^j_0}&=\sum_{k=1}^{n-j}\frac{1}{\rho^{j+1}_0-\rho^j_k}\leq \sum_{k=1}^{n-j}\frac{1}{\frac{\rho^j_0+\rho^j_1}{2}-\rho^j_k}\leq  \sum_{k=1}^{n-j}\frac{1}{k-\frac12}=2H_{2n-2-j}-H_{n-j},
\end{align}
where we have used items (1) and (2) from the proof of theorem \ref{theorem2}. This translates to the more general result 
\begin{align}
    \rho_0^j\leq \rho_0^0-j(2H_{2n-3}-H_{n-1})^{-1}=  \frac12(n-m^2)-r-1-j(2H_{2n-3}-H_{n-1})^{-1},
\end{align}
which is what we wanted.

\begin{figure}[htp]
    \centering
    \includegraphics[width=7.8cm]{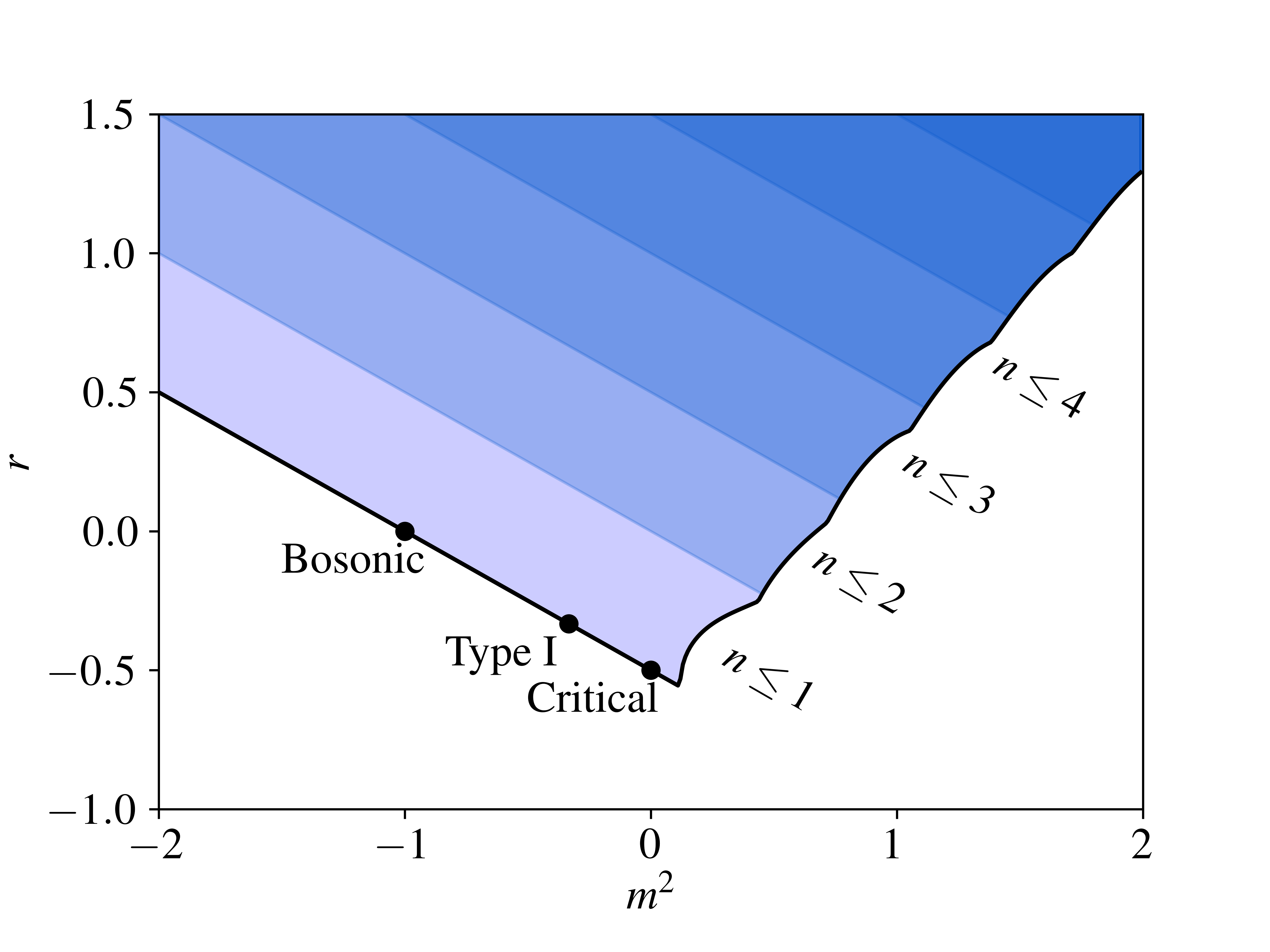}
    \includegraphics[width=7.8cm]{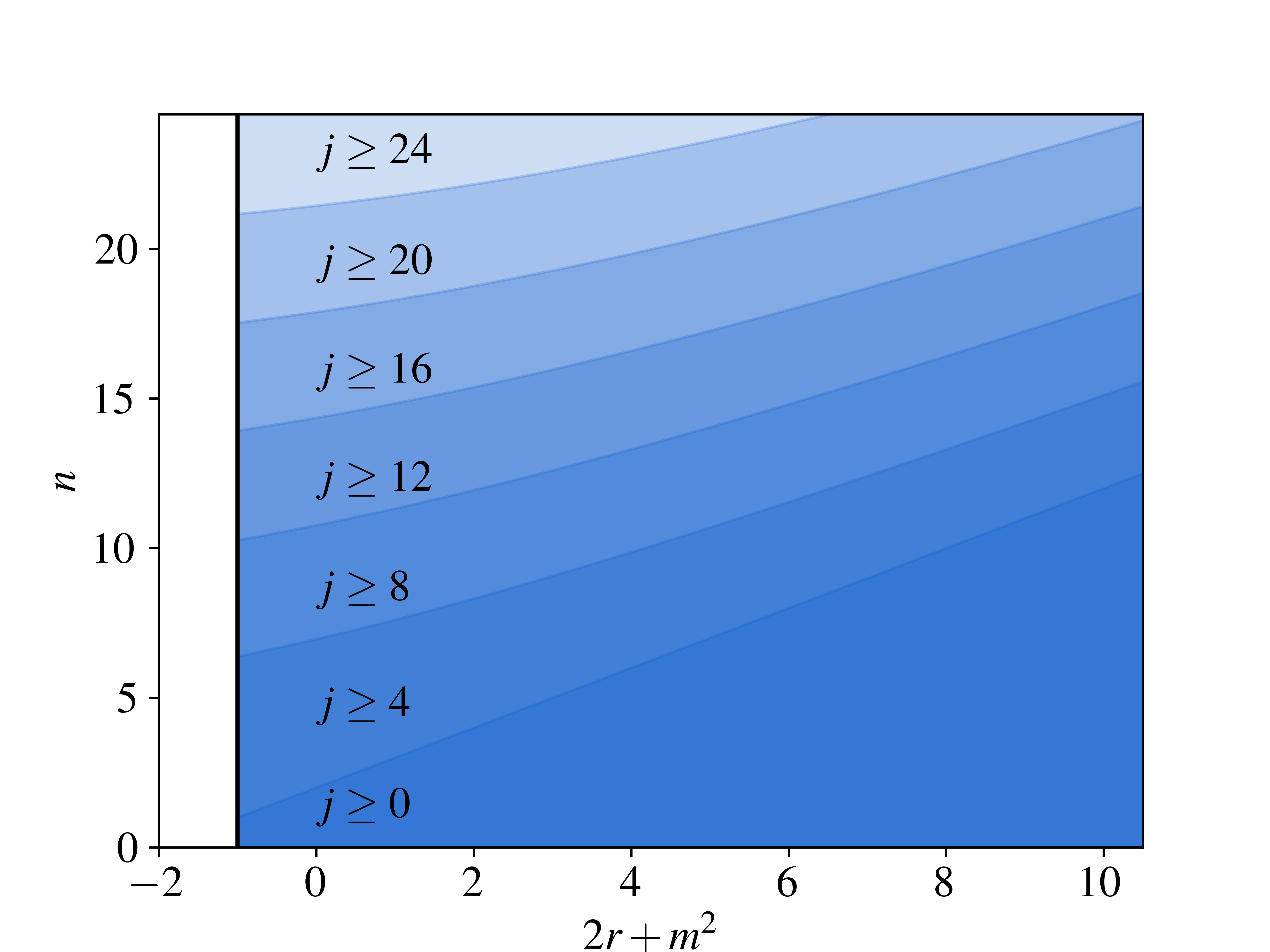}
    \caption{Positivity bounds from theorem \ref{theorem3} for $a=1$. Left: Plots regions in $(m^2,r)$ space. The white region indicates where the hypergeometric amplitude is non-positive for all $D$. The blue regions indicate where theorem \ref{theorem3} guarantees that $B_{n,j}^1\geq 0$ for all $n$ below some threshold value, with no assumption made on the value of $j$. Right: Regions in $(2r+m^2,n)$ space where theorem \ref{theorem3} guarantees $B_{n,j}^1\geq 0$ when $j$ lies above some threshold. The plot has been analytically continued to $n\in\R$ to make the convex-upward trend clearer.}
    \label{redblue}
\end{figure}

\hspace{0pt}

As a corollary to this theorem, we have an argument for positivity at large $r$; more precisely, for any $n$, $j$, and $m^2$, we have shown that there exists a value $r'$ such that $B_{n,j}(r,m^2)\geq 0$ for all $r\geq r'$. This theorem also demonstrates that we expect positivity to hold at large $j$. This is consistent with the conjecture of \cite{Wang:2024wcc} (motivated by the notion of ``low-spin dominance'' introduced in \cite{Bern:2021ppb}) that any non-positive region of parameter space is sculpted out by $B_{n,j}^1$ coefficients exclusively at small values of $j$. Empirically, this condition only seems to hold for coefficients in the region that is not ruled out by theorems \ref{theorem1}, \ref{theorem2}, and conjecture \ref{conjec}; in the regions that are ruled out by these theorems, it is necessary to look for coefficients at high $j$ in order to prove non-unitarity.

\section{Double Contour Representation}
\label{sec:contour}

It is difficult to achieve a positivity bound stronger than theorem \ref{theorem3} via direct analysis of (\ref{eq:gegen}). In this section we present an alternate formula for the partial-wave coefficients of the hypergeometric amplitude at arbitrary $m^2$ and $r$ via a procedure generalizing the work of \cite{Arkani-Hamed:2022gsa} on the Veneziano amplitude. We will see that this formula proves positivity for all coefficients in a large region of parameter space.

\subsection{Derivation}

Throughout the derivation we will suppress all positive constant prefactors that appear in order to prevent the expressions from becoming unmanageably long; these constants will all be restored at the end. We recall the residues of the hypergeometric amplitude are
\begin{align}
    R_n(t)=\frac{(t-m^2+r+1)_{(n-1+a)}}{(r+1)_{(n)}}.
\end{align}
We can re-express the residue as a contour integral via the Cauchy integral formula:
\begin{align}
    R_n(t)&=\frac1{(r+1)_{(n)}}\pdv[n-1+a]{z}\Big|_{z=0}(1-z)^{-t-r-1+m^2}\\
&=\frac{1}{(r+1)_{(n)}}\oint_{z=0}\frac{\dd z}{2\pi i}\frac{(1-z)^{-1-t-r+m^2}}{z^{n+a}}.
\end{align}
Rewrite $t$ in terms of the scattering angle $x=\cos\theta$, and substitute $z$ with a variable $u$ such that $1-z=e^{-u}$, to obtain
\begin{align}
    R_n(t)= \frac{1}{(r+1)_{(n)}}\oint_{u=0}\frac{\dd u}{2\pi i}\frac{\exp[\frac{u}{2}\left((n-3m^2)x+n+2r+m^2+2a\right)]}{(e^u-1)^{n+a}}.\label{eq:residue_contour}
\end{align}
To extract the $j$th Gegenbauer coefficient, use the Rodrigues formula and integrate by parts:
\begin{align}
    B_{n,j}&= \int_{-1}^1\dd x\>(1-x^2)^{J}\dv[j]{x}R_n(t)
\end{align}
where we have defined $J= (D-4)/2+j$. The $j$th derivative of (\ref{eq:residue_contour}) can be computed explicitly, yielding
\begin{align}\
    B_{n,j}= \frac{1}{(r+1)_{(n)}}\oint_{u=0}\frac{\dd u}{2\pi i}\frac{u^j\exp[u\left(\frac{n+m^2}{2}+a+r\right)]}{(e^u-1)^{n+a}}\int_{-1}^1\dd x\>(1-x^2)^{J}e^{xu(n-3m^2)/2}.\label{hard}
\end{align}
The integral over $x$ in (\ref{hard}) does not have a neat closed form, but following the procedure of \cite{Arkani-Hamed:2022gsa}, there is a useful way it can be rewritten. The goal will be to rewrite it as an expression involving a partial derivative with respect to $u$, and then to re-express that partial derivative as a contour integral via the Cauchy integral formula. The key step in doing this begins by recognizing that if we restrict to $D\in 2\Z$, then we can expand the integrand as a derivative via the following identity:
\begin{align}\begin{split}
    (1-x^2)^{J}e^{\beta x}&= (- x\pm 1)^{J} e^{\mp \beta
     }\pdv[J]{\beta}e^{\beta ( x\pm 1)}.
     \end{split}
\end{align}
Commuting this derivative through the integral and manipulating the bounds of integration, we get the following form:
\begin{align}
    \int_{-1}^1\dd x\>(1-x^2)^{J}e^{\beta x}=e^{\beta
    } \pdv[J]{\beta}\int_{-1}^\infty  (- x-1)^{J} e^{\beta (x-1)}\dd x -e^{ -\beta
    } \pdv[J]{\beta} \int_1^\infty  (- x+1)^{J} e^{\beta (x+1)}\dd x.
\end{align}
We change variables $y=x\pm 1$ so that both integrals become identical, leaving the formula
\begin{align}
   \int_{-1}^1\dd x\>(1-x^2)^{J}e^{\beta x}=(-1)^J\left[e^{\beta
    } \pdv[J]{\beta}e^{-2\beta}\int_0^\infty   y^{J}e^{\beta y}\dd y - e^{-\beta
    } \pdv[J]{\beta} e^{2\beta}\int_0^\infty   y^{J}e^{\beta y}\dd y\right],
\end{align}
and the remaining integral here is the Laplace transform of $y^J$. Computing it yields the final result:
\begin{align}
    \int_{-1}^1\dd x\>(1-x^2)^{J}e^{\beta x}=e^{-\beta
    } \pdv[J]{\beta} \frac{e^{2\beta}}{\beta ^{J+1}}-e^{\beta
    } \pdv[J]{\beta}\frac{e^{-2\beta}}{\beta ^{J+1}}.
\end{align}
We can plug this back into (\ref{hard}) in order to obtain
\begin{align}\begin{split}
    B_{n,j}= &\frac{1}{(r+1)_{(n)}}\oint_{u=0}\frac{\dd u}{2\pi i}\frac{u^je^{u\left(2m^2+a+r\right)}}{(e^u-1)^{n+a}} \pdv[J]{u} \frac{e^{u(n-3m^2)}}{u^{J+1}}
    - \frac{u^je^{u\left(n-m^2+a+r\right)}}{(e^u-1)^{n+a}} \pdv[J]{u} \frac{e^{-u(n-3m^2)}}{u^{J+1}}.
\end{split}\end{align}
We can simplify this by a change of variables $u\to -u$ in the second term, which leaves the shorter expression
\begin{align}
    B_{n,j}= &\frac{1}{(r+1)_{(n)}}\oint_{u=0}\frac{\dd u}{2\pi i}\frac{u^j(e^{u(2m^2+a+r)}-(-1)^{n+j+a}e^{u(m^2-r)})}{(e^u-1)^{n+a}}\pdv[J]{u}\frac{e^{u(n-3m^2)}}{u^{J+1}}.
\end{align}
Integrate by parts $J$ times and then rewrite the derivative with respect to $u$ using the Cauchy integral formula:
\begin{align}
    B_{n,j}= &\frac{1}{(r+1)_{(n)}}\oint_{u=0}\frac{\dd u}{2\pi i}\frac{e^{u(n-3m^2)}}{u^{J+1}}\oint_{v=0}\frac{\dd v}{2\pi i}\frac{e^{(u-v)(2m^2+a+r)}-(-1)^{n+j+a}e^{(u-v)(m^2-r)}}{(u-v)^{-j}v^{J+1}(e^{u-v}-1)^{n+a}}.
\end{align}
After some algebra, the final result is
\begin{tcolorbox}[
    title= \textbf{Hypergeometric Contour Formula}]

\begin{align*}
    B_{n,j}^a(r,m^2,D)=\frac{c^a_{n,j,D,m^2}} {(r+1)_{(n)}}\oint_{u=0}\frac{\dd u}{2\pi i}\oint_{v=0}\frac{\dd v}{2\pi i}\left[\frac{(v-u)^j e^{v(2m^2+r+a)+u(m^2-r)}}{(uv)^{\frac{D-2}{2}+j}(e^{v}-e^{u})^{n+a}}
    -(u\leftrightarrow v)\right]
\end{align*}
\end{tcolorbox}
We have also restored the overall normalization, which is given by
\begin{align}
    c^a_{n,j,D,m^2}=(n-1+a)!\frac{2^{D-3}\Gamma(\frac{D-3}{2})(j+\frac{D-3}{2})(j+\frac{D-4}{2})!}{\sqrt \pi\left(n-3m^2\right)^{D+j-3}}\label{eq:normalization}
\end{align}
and is manifestly positive. We note that the contour formula only holds when $D\in 2\Z$. Since positivity at a given $D$ 
implies positivity at all smaller $D$, it will be useful to write results in terms of the largest even integer smaller than $D$, which we denote $\tilde D=2\lfloor \frac D2\rfloor$. 

In the case where $r=0$, this reduces to the analogous contour representation developed by \cite{Arkani-Hamed:2022gsa} for the type-I and bosonic Veneziano amplitudes at $(m^2,a)=(0,0)$ and $(-1,1)$ respectively.\footnote{The $(r,m^2,a)=(0,-1,1)$ case of our expression differs slightly from the analogous equation 1.18 of \cite{Arkani-Hamed:2022gsa}, but this is only due to conventions. We use $n=0,1,\ldots$, while in the bosonic case \cite{Arkani-Hamed:2022gsa} defines $n=-1,0,1,\ldots$} We also see that this formula retains the invariance under the change $a=0\to a=1$, $n\to n-1$, $r\to r-\frac13$, $m^2\to m^2-\frac13$ that was outlined in (\ref{eq:inv}). Because this symmetry allows us to easily transpose results between the bosonic and type-I versions of the hypergeometric amplitude, we will fix $a=1$ for the remainder of the paper for the sake of simplicity.

\subsection{A Positive Region}
\label{sec:positive-region}
\begin{figure}[htp]
    \centering
    \includegraphics[width=14cm]{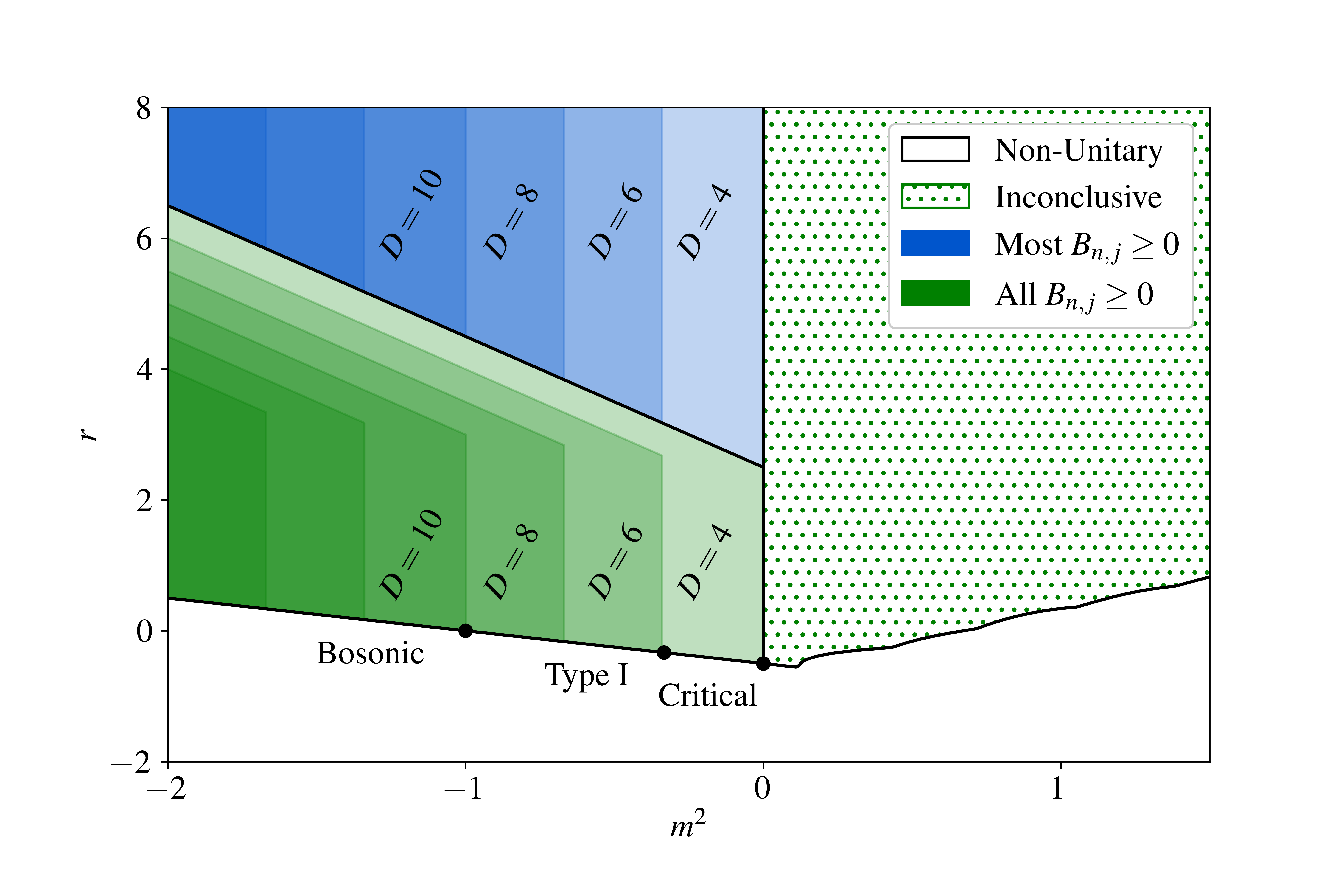}
    \includegraphics[width=14cm]{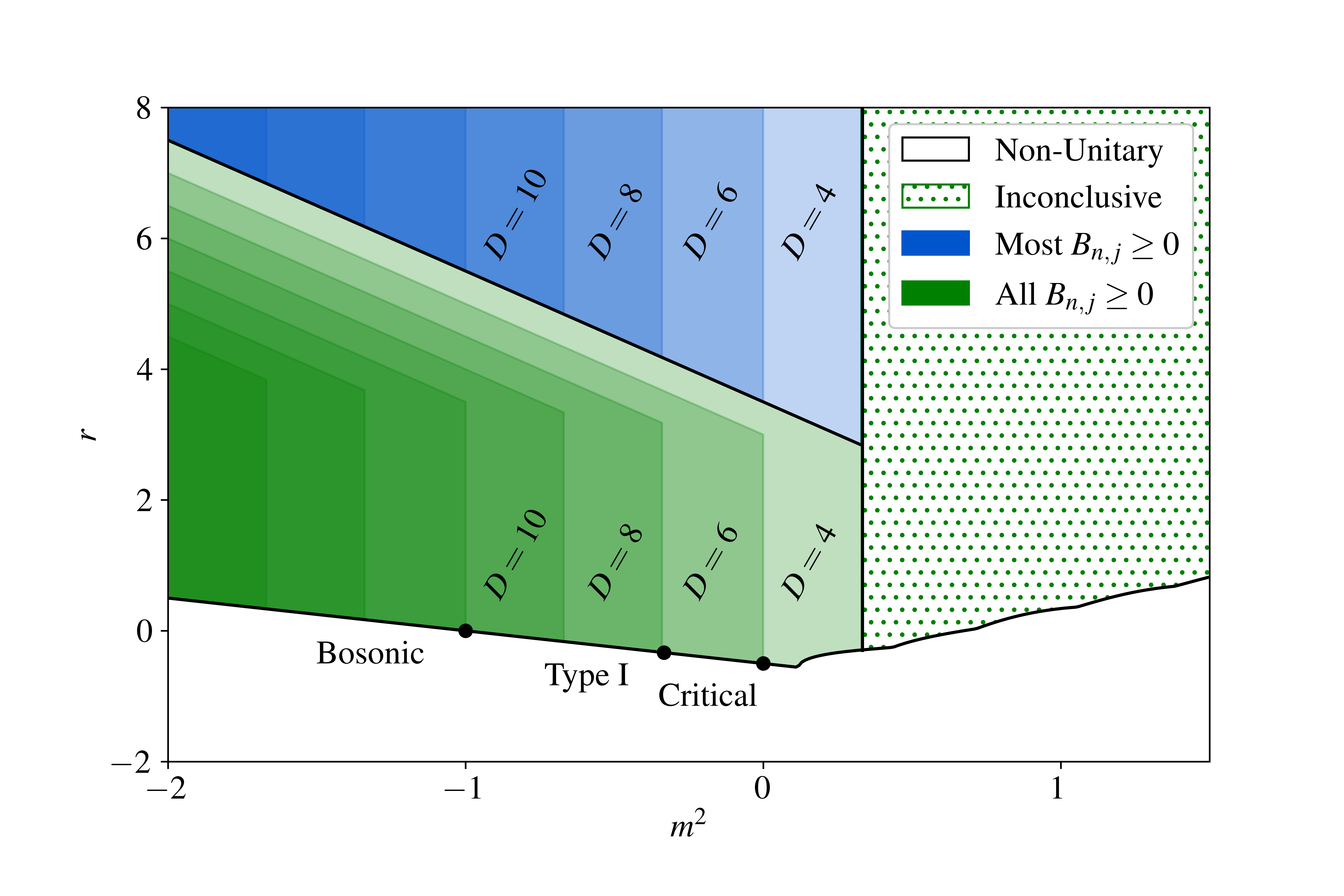}
    \caption{Positivity bounds given for all coefficients (top) by theorem \ref{theorem4} and coefficients with $n-j$ odd (bottom) by theorem \ref{theorem5}. The blue regions depict where $B_{n,j}\geq 0$ whenever $n-j\geq 2^k\geq 2m^2+r+(\tilde D-6)/4$ for some $k\in\mathbb N$. The green regions denote where theorems \ref{theorem4} and \ref{theorem5}, combined with manual computation of leading Regge trajectories up to $n-j\leq 4$, guarantee that $B_{n,j}\geq 0$ for all $n$ and $j$ (top) and all odd $n$ and $j$ (bottom). The boundary between the green and blue regions can be pushed arbitrarily high with a finite amount of work.}
    \label{fig:odd}\label{fig:green}
\end{figure}

The contour formula can be used to demonstrate positivity of all coefficients in a large region of $m^2-r$ space. For simplicity we will fix $a=1$ for the remainder of this section, as the analogous results for $a=0$ can be obtained straightforwardly via (\ref{eq:inv}).
\begin{theorem}\label{theorem4}
    $B_{n,j}^1\geq 0$ when $2r+m^2+1\geq 0$, $m^2\leq \frac{4-\tilde D}{6}$, and there exists some $k\in\mathbb N$ such that $n-j+1\geq 2^k\geq 2m^2+r+\frac{\tilde D-2}{4}$. Here $\tilde D=2\lfloor \frac D2\rfloor$. This bound is depicted in Fig. \ref{fig:green}.
\end{theorem}

Before proceeding with the proof, it is worth giving some interpretation of this result. This theorem establishes a region where all but a finite number of Regge trajectories at low $n-j$ are positive.\footnote{It is tempting to defer to theorem \ref{theorem3} here to prove positivity at large $j$, but this unfortunately isn't quite strong enough.} However, it is also possible to manually verify positivity of all $B_{n,j}^1$ coefficients at a given $n-j$; the coefficients are computed up to $n-j=6$ in Appendix \ref{sec:regge}. Therefore, when supplemented with a finite number of tests at low $n-j$, theorem \ref{theorem4} can be used to establish an arbitrarily large region where \textit{all} $B_{n,j}^1$ coefficients are positive. This region is depicted in Fig. \ref{fig:green}.

\hspace{0pt}

\textit{Proof}. The constraints given by the theorem imply that the prefactor $1/(r+1)_{(n)}$ in the contour formula is positive, so we may ignore its sign. Define new variables $u=\log(1-x)$, $v=\log(1-y)$. Then for any $\gamma\in\Z$, we may recast the contour integral into the following form:
\begin{align}\begin{split}
    \oint_{x=0}\frac{\dd x}{2\pi i}\oint_{y=0}\frac{\dd y}{2\pi i}&I^\pm_{\alpha, \beta,\gamma}(x,y)\left(\frac{(1-x)^{-1}(1-y)^{-1}}{(\log(1-x)\log(1-y))^{2}}\right)^{\frac{D-2}{4}}\frac{\left(\frac{1}{\log(1-x)}-\frac1{\log(1-y)}\right)^j}{(x-y)^{n+1-\gamma}},\label{eq:transform}
\end{split}
\end{align}
where $\alpha=2m^2+r+\frac{D-2}{4}$, $\beta= r-m^2-\frac{D-6}{4}$, and we have defined the function
\begin{align}
    I^{\pm}_{\alpha,\beta,\gamma}(x,y)=\left[\frac{(1-y)^{\alpha}}{(1-x)^{\beta}}\pm \frac{(1-x)^{\alpha}}{(1-y)^{\beta}}\right]\frac{1}{(x-y)^{\gamma}}.
\end{align}
We use $I^+$ when $n-j$ is even and $I^-$ when $n-j$ is odd. 

\hspace{0pt}

We will say a meromorphic function $f(z)$ is \textit{positive} if its Laurent expansion about $z=0$ has exclusively non-negative coefficients. We will also say that a meromorphic function of two variables $f(x,y)$ is positive if its Laurent expansion, first about $y=0$ and then about $x=0$, has exclusively non-negative coefficients. The residue of a positive function at zero will always be non-negative, and any sum or product of two positive functions will be positive. Therefore it suffices to show that the integrand of (\ref{eq:transform}) can be written as a product of positive functions.

\hspace{0pt}

First, the function $\left(\frac{1}{\log(1-x)}-\frac1{\log(1-y)}\right)^j/(x-y)^{n+1-\gamma}$ is positive for any integer $\gamma\leq n-j+1$ -- this is proven in Appendix \ref{sec:lemma2}. The second factor is also positive; to see this, it suffices to prove that
\begin{align}
    f(z)=-\frac{(1-z)^{-\frac12}}{\log(1-z)}\label{eq:logpower}
\end{align}
is positive, as the second factor in (\ref{eq:transform}) is built from products of $f(z)$. We can equivalently write (\ref{eq:logpower}) as the integral
\begin{align}
    f(z)=\frac1z\int_0^1(1-z)^{s-\frac12}\dd s.
\end{align}
From this form, we can compute the $m$th Laurent coefficient of $f$:
\begin{align}
    \pdv[m+1]{z}\Big|_{z=0}(zf(z))=(m+1)\int_0^1\left(\frac12-s\right)_{(m)}\dd s.
\end{align}
The integrand is positive for all $s<1/2$, and it is dominated by the left half of the integration domain, so $f(z)$ is positive. From here, it remains to prove that if $-\beta\leq \alpha\leq \beta$, then $I_{\alpha,\beta,\gamma}^+(x,y)$ and $I_{\alpha,\beta,\gamma}^-(x,y)$ are both positive for some $\gamma\leq n-j+1$. We break this into two cases:

\hspace{0pt}

\textbf{Case 1: $\alpha\leq 0$}

Take $\gamma=0$. Then positivity of $I^+_{\alpha,\beta,0}(x,y)$ immediately follows from the positivity of the function $f(x)=(1-x)^{\rho}$ for all $\rho\leq 0$. Showing positivity of $I^-_{\alpha,\beta,\gamma}$ is a little more involved: we must take $\gamma=1$ and write
\begin{align}
    I^-_{\alpha,\beta,1}(x,y)=\left[\frac{(1-y)^{\alpha}}{(1-x)^{\beta}}-\frac{(1-x)^{\alpha}}{(1-y)^{\beta}}\right]\frac{1}{x-y}.\label{eq:g}
\end{align}
Then we expand (\ref{eq:g}) using the generalized binomial theorem to obtain 
\begin{align}
    I^-_{\alpha,\beta,1}(x,y)=\sum_{k=0}^\infty\sum_{l=0}^\infty\binom{-\alpha+k-1}{k}\binom{\beta+\ell-1}{\ell}\frac{y^k x^\ell-x^k y^\ell}{x-y}
\end{align}
where the binomial coefficients are analytically continued to values in $\R$. If we rearrange the sum so that $k<\ell$,
\begin{align}
    I_{\alpha,\beta,1}^-(x,y)=\sum_{l=0}^\infty \sum_{k=0}^{\ell-1}\left[\binom{-\alpha+k-1}{k}\binom{\beta+\ell-1}{\ell}-\binom{\beta+k-1}{k}\binom{-\alpha+\ell-1}{\ell}\right]\frac{y^k x^\ell-x^k y^\ell}{x-y},  \label{bigeq}
\end{align}
then $(y^kx^\ell-x^ky^\ell)/(x-y)$ is a homogeneous polynomial with positive coefficients,
\begin{align}
    \frac{y^kx^\ell-x^ky^\ell}{x-y}=\sum_{m=0}^{\ell-k-1} x^{k+m}y^{\ell-1-m},
\end{align}
and the factor in brackets in (\ref{bigeq}) is positive under our constraints $k< \ell$ and $-\beta\leq \alpha\leq 0$. 

\hspace{0pt}

\textbf{Case 2: $\alpha\geq 0$}

First we consider $0\leq \alpha\leq 1$ and choose $\gamma=1$. Write $I^-_{\alpha,\beta,1}(x,y)$ as
\begin{align}
    I_{\alpha,\beta,1}^-(x,y)=\frac{1}{(1-y)^{1-\alpha}(1-x)^{1-\alpha}}\times\left[\frac{1-y}{(1-x)^{\alpha+\beta+1}}-\frac{1-x}{(1-y)^{\alpha+\beta+1}}\right]\frac{1}{x-y}.
\end{align}
The first factor is a positive function, and computing the Taylor series of the second factor yields
\begin{align}\begin{split}
    \sum_{k,\ell=0}^\infty x^\ell y^k \binom{\alpha+\beta+\ell+k}{\alpha+\beta}\Bigg[\frac{\alpha+\beta}{k+\ell+1}+\delta(k)+\delta(\ell)\Bigg].\label{bigcoefficient}\end{split}
\end{align}
This expression is manifestly positive, so we have obtained positivity of $I^-_{\alpha,\beta,1}(x,y)$ for $0\leq \alpha\leq 1$. Next, some algebra shows that
\begin{align}
    I^+_{2\alpha,2\beta,2\gamma}(x,y)&=I^-_{\alpha,\beta,\gamma}(x,y)^2+I^+_{\alpha-\beta,\beta-\alpha,2\gamma}(x,y).\label{eq:Iplus}
\end{align}
Since we have already shown that both terms on the right hand side are positive when $0\leq\alpha\leq 1$, $\alpha\leq \gamma$ and $\alpha\leq \beta$, (\ref{eq:Iplus}) guarantees positivity of $I^+_{\alpha,\beta,\gamma}(x,y)$ when $0\leq \alpha\leq 2$, $\alpha\leq \gamma$, and $\alpha\leq \beta$. More generally, (\ref{eq:Iplus}) shows that for any $\alpha\geq 0$, positivity of $I^-_{\alpha,\beta,\gamma}$ implies positivity of $I^+_{2\alpha,2\beta,2\gamma}(x,y)$ when $\alpha\leq \beta$. Some more algebra can be used to show that
\begin{align}
    I^-_{2\alpha,2\beta,2\gamma}(x,y)&=I^+_{\alpha,\beta,\gamma}(x,y)I^-_{\alpha,\beta,\gamma}(x,y),
\end{align}
which means that positivity of $I^\pm_{\alpha,\beta,\gamma}(x,y)$ implies positivity of $I^-_{2\alpha,2\beta,2\gamma}(x,y)$. Repeating this procedure, we find for arbitrary $\alpha,\beta,\gamma$ that if $0\leq \alpha\leq\beta$ and $\alpha\leq 2^k\leq \gamma$ for some $k\in \mathbb N$, then 
positivity of $I_{\alpha,\beta,\gamma}^\pm(x,y)$ follows from positivity of $I_{ 2^{-k}\alpha, 2^{-k}\beta,1}^-(x,y)$. Given that (\ref{bigcoefficient}) is a positive function for $0\leq \alpha \leq 1$, choosing $k$ such that $2^{-k}\alpha\leq 1$ completes the proof.

\subsection{Odd Coefficients}
\label{sec:even}
If we restrict our attention to the case where $n-j$ is odd, there is a larger region where we can prove that $B_{n,j}^1$ becomes positive.

\begin{theorem}\label{theorem5}
    $B_{n,j}^1\geq 0$ when $2r+m^2+1\geq 0$, $n-j$ is odd, $m^2\leq \frac{6-\tilde D}{6}$, and $n-j\geq 2^k\geq 2m^2+r+\frac{\tilde D-2}{4}$ for some $k\in\mathbb N$.
\end{theorem}
\textit{Proof.} To prove the statement, it suffices to show that for all $-\beta\leq \alpha\leq \beta+1$ there is a $\gamma$ such that $I^-_{\alpha,\beta,\gamma}(x,y)$ is positive. Theorem \ref{theorem4} has already proven this for $-\beta\leq \alpha\leq \beta$, so it only remains to prove the case $\beta< \alpha\leq \beta+1$. We write
\begin{align}
    I^-_{\alpha,\beta,\gamma}(x,y)&=\frac{(1-y)^{\alpha+\beta}- (1-x)^{\alpha+\beta}}{(x-y)^{\gamma}(1-x)^{\beta}(1-y)^{\beta}}.\label{eq:Iabc}
\end{align}
To begin, write $\alpha+\beta=pq$ for some $p\in\N$, $q\in[0,1]$ ($p$ and $q$ are not unique here, but any choice will do). This allows us to factor (\ref{eq:Iabc}) as
\begin{align}
    I^-_{\alpha,\beta,\gamma}(x,y)&=\frac{(1-y)^{q}- (1-x)^{q}}{x-y} \sum_{k=0}^{p-1}\frac{(1-y)^{kq-\beta}(1-x)^{(p-k-1)q-\beta}}{(x-y)^{\gamma-1}}.
\end{align}
We can recast this as a relation on $I^\pm_{\alpha,\beta,\gamma}$:
\begin{align}
    I^-_{\alpha,\beta,\gamma}(x,y)&=I^-_{q,0,1}(x,y)\sum_{k=(p-1)/2}^{p-1}I^+_{kq-\beta,\beta-(p-k-1)q,\gamma-1}(x,y)\label{eq:Irecur}
\end{align}
(for simplicity we have assumed $p$ is odd; the formula for even $p$ differs by a single positive term). The factor of $I_{q,0,1}^-(x,y)$ is positive for any $q\in[0,1]$. As long as $\alpha\leq \beta+1$ and $\gamma-1\geq 2^k\geq \alpha$ for some $k\in\mathbb N$, every term in the sum will also be positive by theorem \ref{theorem4}. Therefore $I^-_{\alpha,\beta,\gamma}$ is positive for all $-\beta\leq \alpha\leq \beta+1$.

\hspace{0pt}

It is worth noting that when $2r+m^2\in\N$, we can choose $q=1$, causing the $I_{q,0,1}^-(x,y)$ prefactor in (\ref{eq:Irecur}) to vanish. Then (\ref{eq:Irecur}) translates to a curious relation between the partial wave coefficients of the bosonic and type-I amplitude's residues. When $n-j$ is odd, we have
\begin{align}
    B^1_{n,j}(r,m^2,D)&=\frac1n\sum_{k=0}^{r+m^2/2}\frac{(k-\frac{m^2}{2}+\frac 23)_{(n-1)}}{(r+1)_{(n)}}B^1_{n-1,j}\left(k-\frac{m^2}2-\frac13,m^2-\frac13,D\right)\\
    &=\frac1n\sum_{k=0}^{r+m^2/2}\frac{(k-\frac{m^2}{2}+1)_{(n)}}{(r+1)_{(n)}} B^0_{n,j}\left(k-\frac{m^2}2,m^2,D\right),
\end{align}
and when $n-j$ is even, a similar argument yields 
\begin{align}
    B^1_{n,j}(r,m^2,D)&=\frac1n\sum_{k=0}^{r+m^2/2}(-1)^k\frac{(k-\frac{m^2}{2}+1)_{(n)}}{(r+1)_{(n)}}B^0_{n,j}\left(k-\frac{m^2}2,m^2,D\right).
\end{align}
It is also possible to directly derive these relations from (\ref{eq:gegen}).

\subsection{Even Coefficients}
\label{sec:odd}

It is also possible to improve the positive region for even $n+j$ beyond the bound given in theorem \ref{theorem4}, although the improvement is not as significant as what was possible for odd $n+j$. In this section, we will prove two theorems:

\begin{theorem}
$B_{n,j}^1\geq 0$ when $n+j$ is even, 
 $D= 4$, and $2m^2+r\leq -\frac{1}{\sqrt 6}$.\label{theorem6}
\end{theorem}

\begin{theorem}
    $B_{n,j}^1\geq 0$ when $n+j$ is even, $D\geq 6$, and the following two conditions hold: 
    \begin{align*}
        2m^2+r+\frac{\tilde D}{4}\leq 1
    \end{align*}
    and
    \begin{align*}
        \frac{j+1}{n+1}&\leq \frac{G_{\tilde D-4}}{G_{\tilde D-5} \left(2m^2+r+\frac{D-2}{4}\right)}-1\label{GConstraint}
    \end{align*}
    where $G_m$ denotes the absolute value of the $m$th Gregory coefficient, which is defined from the Laurent series
    \begin{align}
        \frac{z}{\log(1-z)}=\sum_{m=0}^\infty G_m z^m
    \end{align}\label{theorem7}
\end{theorem}

\textit{Proof.}
To begin, we define the following Laurent series:
\begin{align}
    \frac{(1-z)^p}{(-\log(1-z))^m}&=\sum_{k=-m}^\infty a_k(p,m) z^k\\
    \frac{\left(\frac{1}{\log(1-x)}-\frac1{\log(1-y)}\right)^j}{(x-y)^n}&=\sum_{\alpha,\beta}b_{\alpha,\beta}(n,j)x^\alpha y^\beta
\end{align}
where $m$, $n$, and $j$ are integers, but $p$ can be any real number. With this expansion, we can rewrite the integrand in the contour formula as
\begin{align}
    B_{n,j}^1=\frac{c_{n,j,D,m^2}^1}{(r+1)_{(n+1)}}\oint_{x=0}\frac{\dd x}{2\pi i}\oint_{y=0}\frac{\dd y}{2\pi i}\left(\sum_{k,\ell=-q}^\infty  (a_k\tilde a_\ell+\tilde a_k a_\ell) y^kx^\ell\right)\left(\sum_{\alpha,\beta}b_{\alpha,\beta}x^\alpha y^\beta\right)
\end{align}
where to make this more concise we have defined $q= (D-2)/2$, $a_k= a_k(2m^2+r,q)$, $\tilde a_k = a_k(1+m^2-r,q)$, and $b_{\alpha\beta}= b_{\alpha\beta}(n,j)$. If we take the Cauchy product of these series, we find that the coefficient on the $(xy)^{-1}$ term evaluates to the finite sum
\begin{align}
    B_{n,j}^1=\frac{c_{n,j,D,m^2}^1}{(r+1)_{(n)}}\sum_{\alpha=-q+1}^{n+j+q-1}\sum_{\beta=-q+1}^{j}(a_{\alpha-1}\tilde a_{\beta-1}+ \tilde a_{\alpha-1} a_{\beta-1})b_{-\alpha,-\beta}.\label{eq:cauchy}
\end{align}
Next we state some key inequalities satisfied by the $a_k$, $\tilde a_k$, and $b_{\alpha\beta}$ coefficients that will be useful. They are proved in Appendix \ref{sec:appendix}.
\begin{lemma}
    When $k\geq-q+2$ and one of the following is true,
    \begin{enumerate}
        \item $q=1$ and $p\leq -\frac1{\sqrt 6}$
        \item $q=2$ and $p\leq \frac{-3 - \sqrt 3}{6}$
        \item $q\geq 3$ and $ p\leq \frac{1-q}2$
    \end{enumerate}
    then $a_k\geq 0$ and $2a_{k+1}\geq a_k$.
\end{lemma}

\begin{lemma}
    When $\alpha+\beta+j+n\neq 0$, the $b_{\alpha\beta}$ coefficients satisfy
\begin{align*}
    \frac{n-1+\beta}{j+\beta}\frac{G_{\lceil(\alpha+n)/j\rceil+2}}{G_{\lceil(\alpha+n)/j\rceil+1}}b_{\alpha,\beta+1}\geq \frac{G_{\lceil(\alpha+n)/j\rceil+2}}{G_{\lceil(\alpha+n)/j\rceil+1}}b_{\alpha+1,\beta}\geq b_{\alpha,\beta}\geq 0.
\end{align*}

\end{lemma}
 One can also verify that $b_{\alpha\beta}=0$ when $\alpha+\beta+n+j=1$. Now from these lemmas we can see that $a_{-q+1}$ is the only negative coefficient appearing in either Laurent series; every other $a_k$ and $b_{\alpha\beta}$ coefficient is either positive or zero. Then to show that this sum is positive, we will break it into five smaller sums, isolating the terms that contain a factor of $a_{-q+1}$, and demonstrate that each of these terms is positive. We divide up the terms of (\ref{eq:cauchy}) as follows:

\begin{align}\begin{split}
    &\sum_{\alpha=-q+3}^{n+j+q-1}a_{\alpha-1}\left(\tilde a_{-q}b_{-\alpha,q-1}+\tilde a_{-q+1}b_{-\alpha,q-2}\right)\\
    &+ \sum_{\beta=-q+3}^{j}\tilde a_{\beta-1}\left(a_{-q}b_{q-1,-\beta}+a_{-q+1}b_{q-2,-\beta}\right)\\
    &+\left(a_{q+n+j-2}\tilde a_{-q}b_{-n-j-q+1,q-1}+a_{q+n+j-3}\tilde a_{-q+1}b_{-n-j-q+2,q-2}\right)\\
    &+(a_{-q+1}\tilde a_{-q}b_{q-2,q-1}+a_{-q}\tilde a_{-q+1}b_{q-1,q-2}+a_{-q}\tilde a_{-q}b_{q-1,q-1}+a_{-q+1}\tilde a_{-q+1}b_{q-2,q-2})\\
    &+(\rm{other\ positive\ terms})\\
    &+(a\leftrightarrow \tilde a).\label{eq:bigsum}
\end{split}\end{align}
The $(a\leftrightarrow\tilde a)$ indicates that we must also sum over the previous five lines, with $a$ and $\tilde a$ switched. A more intuitive picture of this grouping of terms is given in table \ref{table1}. 

\begin{table}[htp]
    \centering
    \resizebox{8cm}{!}{
        \begin{tabular}{cccccccccccc}
        & \multicolumn{1}{c}{}            &                       &        &  &                          &                       & $\alpha$                 &                       & \multicolumn{1}{l}{}          & \multicolumn{1}{l}{} &                        \\
        & \multicolumn{1}{l|}{}           &                       &        &  &                          &                       &                          &                       & \multicolumn{1}{l}{}          & \multicolumn{1}{l}{} &                        \\
        & \multicolumn{1}{l|}{}           &                       & \begin{rotate}{30}$-q+1$\end{rotate} &  & \begin{rotate}{30}$-q+2$\end{rotate}                   &                       & \begin{rotate}{30}$\ldots $\end{rotate}                &                       & \multicolumn{1}{l}{\begin{rotate}{30}$n+j+q-2$\end{rotate}} & \multicolumn{1}{l}{} & \begin{rotate}{30}$n+j+q-1$\end{rotate}              \\ \cline{2-12} 
        & \multicolumn{1}{l|}{}           &                       &        &  &                          &                       &                          &                       &                               &                      &                        \\ \cline{4-6} \cline{8-8} \cline{10-12} 
        & \multicolumn{1}{l|}{$-q+1$}     & \multicolumn{1}{c|}{} & +      &  & \multicolumn{1}{c|}{$-$} & \multicolumn{1}{c|}{} & \multicolumn{1}{c|}{+}   & \multicolumn{1}{c|}{} & 0                             &                      & \multicolumn{1}{c|}{+} \\
        & \multicolumn{1}{l|}{}           & \multicolumn{1}{c|}{} &        &  & \multicolumn{1}{c|}{}    & \multicolumn{1}{c|}{} & \multicolumn{1}{c|}{}    & \multicolumn{1}{c|}{} &                               &                      & \multicolumn{1}{c|}{}  \\
$\beta$ & \multicolumn{1}{l|}{$-q+2$}     & \multicolumn{1}{c|}{} & $-$    &  & \multicolumn{1}{c|}{+}   & \multicolumn{1}{c|}{} & \multicolumn{1}{c|}{$-$} & \multicolumn{1}{c|}{} & $-$                           &                      & \multicolumn{1}{c|}{0} \\ \cline{4-6} \cline{8-8} \cline{10-12} 
        & \multicolumn{1}{l|}{}           &                       &        &  &                          &                       &                          &                       &                               &                      &                        \\ \cline{4-6}
        & \multicolumn{1}{l|}{$\ldots j$} & \multicolumn{1}{c|}{} & +      &  & \multicolumn{1}{c|}{$-$} &                       & +                        &                       & 0                             &                      & 0                   \\ \cline{4-6}
\end{tabular}}
    \caption{Decomposition of positive and negative terms in the convolution. Each $+$/$-$/$0$ represents the sign on the term $a_{\alpha-1}\tilde a_{\beta-1}b_{-\alpha,-\beta}$ in (\ref{eq:bigsum}). Our goal is to show that the sum of terms in each box evaluates to a positive number.}
    \label{table1}
\end{table}

\hspace{0pt}

The positivity of the fifth term is already manifest, since all instances of $a_{-q+1}$ are contained in the first four terms. The positivity of the first two terms follows from a direct calculation that $a_{-q}(p,q)=1$ and $a_{-q+1}(p,q)=-p-\frac q2\geq -\frac 12$ by our assumption that $2m^2+r+D/4\leq 1$. This inequality, combined with lemma 2 and the fact that $G_{k+1}/G_k\geq \frac12$ for all $k>1$, straightforwardly gives positivity of the first two terms. Positivity of the fourth term follows from a similar argument, but requires us to place an upper bound on $(j+1)/(n+1)$ when $D\neq 4$. Lastly, we can see positivity of the third term as follows: a direct calculation tells us that for any $s\in\Z$,
\begin{align}
    b_{-n-j-s,s}=\binom{n+s-1}{s+j}.
\end{align}
Using this in combination with our known expressions $a_{-q}(p,q)=1$ and $a_{-q+1}(p,q)=-p-\frac q2$, it suffices to show that
\begin{align}
    a_{q+n+j-2}\binom{n+q-2}{j+q-1}-a_{q+n+j-3}\left(p+\frac q2\right)\binom{n+q-3}{ j+q-2}\geq 0
\end{align}
for $p\leq \frac{1-q}{2}$. Now using that $\binom{n+q-2}{j+q-1}\geq \binom{n+q-3}{ j+q-2}$ and $p+\frac q2\leq \frac12$, it suffices to prove that $a_{q+n+j-2}-a_{q+n+j-3}/2\geq 0$, and this is given by lemma 1. This completes the proof.

\hspace{0pt}

For each $D$ we notice that there is a large region where \textit{all} odd coefficients are positive, regardless of the value of $(j+1)/(n+1)$. This region is given by
\begin{align}
    4m^2+2r+\frac{\tilde D-2}{2}&\leq \frac{G_{\tilde D-4}}{G_{\tilde D-5} }\label{oddconstraint}
\end{align}  
for $D\geq 6$. We also note that there are many ``unused" positive terms left over in (\ref{eq:bigsum}); we expect that one could slightly strengthen the bound (\ref{oddconstraint}) by making use of these terms.

\subsection{Implications for Veneziano Unitarity}

Examining the type-I Veneziano $(r,m^2,a)=(0,0,0)$ and bosonic Veneziano $(r,m^2,a)=(0,-1,1)$ cases, our contour formula reproduces the proof of positivity of the $D\leq 6$ type-I Veneziano amplitude and the $D\leq 10$ bosonic Veneziano amplitude that originally appeared in \cite{Arkani-Hamed:2022gsa}. However, it also gives a result at higher $D$: theorem \ref{theorem5} gives positivity of the type-I Veneziano coefficients when $D=8$ and
\begin{align*}
    \frac{j+1}{n+1}\leq \frac{4}{15}.
\end{align*}
It gives positivity of the bosonic Veneziano coefficients when $D=12$ and
\begin{align*}
    \frac{j+1}{n+1}\leq \frac{13328}{20625}\approx 0.6462.
\end{align*}
A stronger version of lemma 2 could be used to prove positivity of the full amplitudes in $D=8$ and $D=12$ respectively. Unfortunately, these do not get us closer to understanding positivity in $D=10$ and $D=26$, which are more physically significant. 

\hspace{0pt}

Another interesting special case mentioned in \cite{Cheung:2023adk} is the ``critical" amplitude at $(r,m^2,D)=(-1/2,0,5)$. The amplitude is named as such because it exists on a ``cusp" in parameter space: it becomes manifestly non-unitary if we further deform $m^2$ away from $0$, $r<-1/2$, or $D>5$. The contour formula proves positivity of this amplitude in $D=4$, but the $D=5$ case remains out of reach of our methods.

\section{Limiting Cases}
\label{sec:limiting}

\subsection{Positivity at Low Spin}
Another benefit of the double contour formula is that we may fix some $j$ and $D$, and simplify the expression to a point where positivity of all coefficients at arbitrary $n$ becomes manifest. The procedure for doing this is a natural generalization of the method used in \cite{Arkani-Hamed:2022gsa} to demonstrate positivity at low spin for the Veneziano amplitude in $D=10$. This low-spin analysis can also be carried out via a different method in the generalized case of the hypergeometric Coon amplitude as demonstrated in \cite{Rigatos:2024beq, Wang:2024wcc}. The calculations required here quickly become more complicated as we increase $j$ and $D$, so we will only present the simplest case.

\begin{theorem}
    If $D=4$, $r\geq m^2$, and $2r+m^2+1\geq 0$ then $B_{n,0}^1(r,m^2)\geq 0$ for all $n$.
\end{theorem}
\textit{Proof.} Fixing $j=0$, $D=4$, the $B_{n,j}^1$ coefficients are given up to a positive constant by
\begin{align}
    \oint_{x=0}\frac{\dd x}{2\pi i}\oint_{y=0}\frac{\dd y}{2\pi i}&\left[\frac{(1-y)^{2m^2+r}}{(1-x)^{r-m^2+1}}\pm \frac{(1-x)^{2m^2+r}}{(1-y)^{r-m^2+1}}\right]\frac{1}{\log(1-x)\log(1-y)}\frac{1}{(x-y)^{n+1}}.
\end{align}
We may directly evaluate the residue in $y$ to obtain
\begin{align}
    -\oint_{x=0}\frac{\dd x}{2\pi i}\frac{(1-x)^{m^2-r-1}\pm (1-x)^{2m^2+r}}{x^{n+1}\log(1-x)}.
\end{align}
Rewriting the log as an integral and applying the Cauchy integral formula, we see this is equivalent to
\begin{align}
    \pdv[n+1]{x}\Big|_{x=0}\int_0^1 \dd s\>  (1-x)^{m^2-r-1+s}+ (-1)^n(1-x)^{2m^2+r+s}.
\end{align}
We can directly evaluate the derivatives in terms of rising factorials:
\begin{align}
    \int_0^1 \dd s\> (-m^2+r+1-s)_{(n+1)}- (2m^2+r+s-n)_{(n+1)}.
\end{align}
Since $r\geq m^2$, the first term in the integrand is non-negative for all $s\in[0,1]$. Since $2r+m^2+1\geq 0$, the second term has a smaller absolute magnitude than the first. Therefore the integral must be positive.

\subsection{Fixed-Spin Asymptotics}

In this section we establish a large region where $B_{n,j}^1$ becomes positive in the limit as $n\to \infty$ with $j$ held at some fixed value, and another new region where these coefficients are negative. Again, the procedure here generalizes the one presented in \cite{Arkani-Hamed:2022gsa} for the Veneziano amplitude.

\begin{theorem} 
As $n\to \infty$ with $j$ held fixed, the asymptotic relation 
\begin{align}
    B^1_{n,j}\sim \Gamma(r+1)\frac{2^{D-3}\Gamma(\frac{D-3}{2})(j+\frac{D-3}{2})}{\sqrt{\pi}\log(n)^{\frac{D-2}{2}}n^{\frac{D}{2}+2m^2}}\left[\frac{n^{2r+m^2+1}}{\Gamma(r-m^2+1)}+\frac{(-1)^{n-j}}{\Gamma(-r-2m^2)}\right]
\end{align}
holds for $D\in 2\mathbb Z$.

\label{theorem9}
\end{theorem} 

\textit{Proof.} Beginning again with the double contour formula
\begin{align}\begin{split}
    B_{n,j}^1=\frac{c_{n,j,D,m^2}^1}{(r+1)_{(n)}}\oint_{x=0}&\frac{\dd x}{2\pi i}\oint_{y=0}\frac{\dd y}{2\pi i}\left[\frac{(1-y)^{2m^2+r}}{(1-x)^{r-m^2+1}}+(-1)^{n+j}\frac{(1-x)^{2m^2+r}}{(1-y)^{r-m^2+1}}\right]\\&\times\frac{1}{(\log(1-x)\log(1-y))^{\frac{D-2}{2}}}\left(\frac{\frac{1}{\log(1-x)}-\frac1{\log(1-y)}}{x-y}\right)^j(x-y)^{j-n-1},
\end{split}
\end{align}
we can expand the factor $(x-y)^{j-n-1}$ as
\begin{align}
    (x-y)^{-n+j-1}=\sum_{k=0}^\infty \binom{n-j+k}{k}x^{-n+j-1-k}y^{k}.\label{binomexpand}
\end{align}
However, by examining Laurent series of the other factors in the integrand, we find that all terms in (\ref{binomexpand}) with $k> j+\frac{D}{2}-2$ give zero contribution to the residue in $y$. Therefore we may drop all of these terms without altering the value of the integral. Of the terms that remain, the one at $k=j+\frac{D}{2}-2$ dominates in the large $n$ limit; therefore, we substitute the entirety of (\ref{binomexpand}) with this single term:
\begin{align}
    (x-y)^{-n+j-1}\to \binom{n-\frac{D}{2}+2}{j+\frac{D}{2}-2}x^{-n-\frac{D-2}2}y^{j+\frac{D}{2}-2}.
\end{align}
Once we've made this substitution, only the lowest-order Laurent series terms in each of the other factors gives a nonzero contribution to the residue; therefore we can make the following substitutions with no effect on the evaluation of the integral:
\begin{align}\begin{split}
    (-\log(1-y))^{1-D/2}&\to y^{1-D/2}\\\left(\frac{\frac{1}{\log(1-x)}-\frac1{\log(1-y)}}{x-y}\right)^j&\to\frac{1}{(xy)^j}\\(1-y)^\alpha&\to 1
\end{split}\end{align}
Now the integral in $y$ reduces to a simple power law, which we directly evaluate yielding 
\begin{align}\begin{split}
    B_{n,j}^1\sim\frac{c_{n,j,D,m^2}^1}{(r+1)_{(n)}}\frac{n^{j+\frac{D}{2}-2}}{(j+\frac{D}{2}-2)!}\oint_{x=0}&\frac{\dd x}{2\pi i}\frac{x^{-n-j-\frac{D-2}{2}}}{(-\log(1-x))^{\frac{D-2}{2}}}\\&\times\left[(1-x)^{m^2-r-1}+(-1)^{n+j}(1-x)^{2m^2+r}\right].\end{split}
\end{align}
Substitute $x=1+\frac{t}{n}$:
\begin{align}\begin{split}
    B_{n,j}^1\sim\frac{c_{n,j,D,m^2}^1}{(r+1)_{(n)}}\frac{n^{j+\frac{D}{2}-3}}{(j+\frac{D}{2}-2)!}\oint_{t=-n}&\frac{\dd t}{2\pi i}\frac{(1+\frac{t}{n})^{-n-j-\frac{D-2}{2}}}{(-\log(-\frac{t}{n}))^{\frac{D-2}{2}}}\\&\times\left[\left(-\frac{t}{n}\right)^{m^2-r-1}+(-1)^{n+j}\left(-\frac{t}{n}\right)^{2m^2+r}\right].
\end{split}
\end{align}

\begin{figure}[ht]
    \centering
\begin{tikzpicture}[x=0.75pt,y=0.75pt,yscale=-0.75,xscale=0.75]
    
    \draw    (32.5,152.42) -- (296.79,152.42) ;
    \draw    (172.28,49.38) -- (172.28,286) ;
    \draw  [fill={rgb, 255:red, 0; green, 0; blue, 0 }  ,fill opacity=1 ] (78.87,152.5) .. controls (78.87,150.21) and (80.73,148.35) .. (83.02,148.35) .. controls (85.32,148.35) and (87.17,150.21) .. (87.17,152.5) .. controls (87.17,154.8) and (85.32,156.65) .. (83.02,156.65) .. controls (80.73,156.65) and (78.87,154.8) .. (78.87,152.5) -- cycle ;
    \draw  [line width=2.25]  (173,151.9) .. controls (174.22,154) and (175.39,156) .. (176.75,156) .. controls (178.11,156) and (179.28,154) .. (180.5,151.9) .. controls (181.72,149.8) and (182.89,147.8) .. (184.25,147.8) .. controls (185.61,147.8) and (186.78,149.8) .. (188,151.9) .. controls (189.22,154) and (190.39,156) .. (191.75,156) .. controls (193.11,156) and (194.28,154) .. (195.5,151.9) .. controls (196.72,149.8) and (197.89,147.8) .. (199.25,147.8) .. controls (200.61,147.8) and (201.78,149.8) .. (203,151.9) .. controls (204.22,154) and (205.39,156) .. (206.75,156) .. controls (208.11,156) and (209.28,154) .. (210.5,151.9) .. controls (211.72,149.8) and (212.89,147.8) .. (214.25,147.8) .. controls (215.61,147.8) and (216.78,149.8) .. (218,151.9) .. controls (219.22,154) and (220.39,156) .. (221.75,156) .. controls (223.11,156) and (224.28,154) .. (225.5,151.9) .. controls (226.72,149.8) and (227.89,147.8) .. (229.25,147.8) .. controls (230.61,147.8) and (231.78,149.8) .. (233,151.9) .. controls (234.22,154) and (235.39,156) .. (236.75,156) .. controls (238.11,156) and (239.28,154) .. (240.5,151.9) .. controls (241.72,149.8) and (242.89,147.8) .. (244.25,147.8) .. controls (245.61,147.8) and (246.78,149.8) .. (248,151.9) .. controls (249.22,154) and (250.39,156) .. (251.75,156) .. controls (253.11,156) and (254.28,154) .. (255.5,151.9) .. controls (256.72,149.8) and (257.89,147.8) .. (259.25,147.8) .. controls (260.61,147.8) and (261.78,149.8) .. (263,151.9) .. controls (264.22,154) and (265.39,156) .. (266.75,156) .. controls (268.11,156) and (269.28,154) .. (270.5,151.9) .. controls (271.72,149.8) and (272.89,147.8) .. (274.25,147.8) .. controls (275.61,147.8) and (276.78,149.8) .. (278,151.9) .. controls (279.22,154) and (280.39,156) .. (281.75,156) .. controls (283.11,156) and (284.28,154) .. (285.5,151.9) .. controls (286.72,149.8) and (287.89,147.8) .. (289.25,147.8) .. controls (290.61,147.8) and (291.78,149.8) .. (293,151.9) .. controls (293.34,152.48) and (293.67,153.05) .. (294,153.57) ;
    \draw    (349.5,152.42) -- (613.79,152.42) ;
    \draw    (489.28,49.38) -- (489.28,286) ;
    \draw  [fill={rgb, 255:red, 0; green, 0; blue, 0 }  ,fill opacity=1 ] (390.87,152.5) .. controls (390.87,150.21) and (392.73,148.35) .. (395.02,148.35) .. controls (397.32,148.35) and (399.17,150.21) .. (399.17,152.5) .. controls (399.17,154.8) and (397.32,156.65) .. (395.02,156.65) .. controls (392.73,156.65) and (390.87,154.8) .. (390.87,152.5) -- cycle ;
    \draw  [line width=2.25]  (490,151.9) .. controls (491.22,154) and (492.39,156) .. (493.75,156) .. controls (495.11,156) and (496.28,154) .. (497.5,151.9) .. controls (498.72,149.8) and (499.89,147.8) .. (501.25,147.8) .. controls (502.61,147.8) and (503.78,149.8) .. (505,151.9) .. controls (506.22,154) and (507.39,156) .. (508.75,156) .. controls (510.11,156) and (511.28,154) .. (512.5,151.9) .. controls (513.72,149.8) and (514.89,147.8) .. (516.25,147.8) .. controls (517.61,147.8) and (518.78,149.8) .. (520,151.9) .. controls (521.22,154) and (522.39,156) .. (523.75,156) .. controls (525.11,156) and (526.28,154) .. (527.5,151.9) .. controls (528.72,149.8) and (529.89,147.8) .. (531.25,147.8) .. controls (532.61,147.8) and (533.78,149.8) .. (535,151.9) .. controls (536.22,154) and (537.39,156) .. (538.75,156) .. controls (540.11,156) and (541.28,154) .. (542.5,151.9) .. controls (543.72,149.8) and (544.89,147.8) .. (546.25,147.8) .. controls (547.61,147.8) and (548.78,149.8) .. (550,151.9) .. controls (551.22,154) and (552.39,156) .. (553.75,156) .. controls (555.11,156) and (556.28,154) .. (557.5,151.9) .. controls (558.72,149.8) and (559.89,147.8) .. (561.25,147.8) .. controls (562.61,147.8) and (563.78,149.8) .. (565,151.9) .. controls (566.22,154) and (567.39,156) .. (568.75,156) .. controls (570.11,156) and (571.28,154) .. (572.5,151.9) .. controls (573.72,149.8) and (574.89,147.8) .. (576.25,147.8) .. controls (577.61,147.8) and (578.78,149.8) .. (580,151.9) .. controls (581.22,154) and (582.39,156) .. (583.75,156) .. controls (585.11,156) and (586.28,154) .. (587.5,151.9) .. controls (588.72,149.8) and (589.89,147.8) .. (591.25,147.8) .. controls (592.61,147.8) and (593.78,149.8) .. (595,151.9) .. controls (596.22,154) and (597.39,156) .. (598.75,156) .. controls (600.11,156) and (601.28,154) .. (602.5,151.9) .. controls (603.72,149.8) and (604.89,147.8) .. (606.25,147.8) .. controls (607.61,147.8) and (608.78,149.8) .. (610,151.9) .. controls (610.34,152.48) and (610.67,153.05) .. (611,153.57) ;
    \draw   (50.53,152.5) .. controls (50.53,134.56) and (65.08,120.01) .. (83.02,120.01) .. controls (100.97,120.01) and (115.51,134.56) .. (115.51,152.5) .. controls (115.51,170.45) and (100.97,184.99) .. (83.02,184.99) .. controls (65.08,184.99) and (50.53,170.45) .. (50.53,152.5) -- cycle ;
    \draw    (488,137) -- (609,137) ;
    \draw    (489.28,167.69) -- (610.28,167.69) ;
    \draw  [draw opacity=0] (489.28,167.69) .. controls (489.16,167.69) and (489.04,167.69) .. (488.92,167.69) .. controls (480.44,167.69) and (473.57,160.82) .. (473.57,152.35) .. controls (473.57,143.87) and (480.44,137) .. (488.92,137) .. controls (488.98,137) and (489.03,137) .. (489.09,137) -- (488.92,152.35) -- cycle ; \draw   (489.28,167.69) .. controls (489.16,167.69) and (489.04,167.69) .. (488.92,167.69) .. controls (480.44,167.69) and (473.57,160.82) .. (473.57,152.35) .. controls (473.57,143.87) and (480.44,137) .. (488.92,137) .. controls (488.98,137) and (489.03,137) .. (489.09,137) ;  
    \draw   (98.31,172.67) -- (112.77,168.47) -- (106.52,182.17) ;
    \draw   (581.5,130.94) -- (595.43,136.67) -- (582.01,143.49) ;
    
    \draw (312.14,149) node [anchor=north west][inner sep=0.75pt]    {$\rightarrow $};
    \draw (81.99,129.84) node [anchor=north west][inner sep=0.75pt]    {$-n$};
    \draw (393.99,129.84) node [anchor=north west][inner sep=0.75pt]    {$-n$};

    \end{tikzpicture}

    \caption{Hankel contour deformation.}
        \label{fig:hankel}

\end{figure}
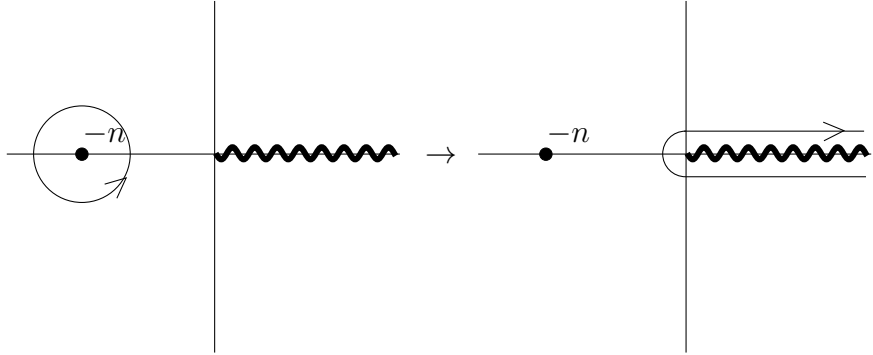

We can deform the contour encircling $t=-n$ to the Hankel contour $\mc H$ encircling the branch cut along the positive real axis as depicted in Fig. \ref{fig:hankel}. The contribution from the arc at infinity vanishes, and the choice of contour no longer depends on $n$. Now as we send $n\to\infty$, over any bounded subset of $\mathbb C$ the integrand uniformly asymptotes to
\begin{align}
    B_{n,j}^1\sim\frac{c_{n,j,D,m^2}^1}{(r+1)_{(n)}}\frac{n^{j+\frac{D}{2}-3}}{(j+\frac{D}{2}-2)!\log(n)^{\frac{D-2}{2}}}\oint_{\mathcal H}\frac{\dd t}{2\pi i}\left[\left(-\frac{t}{n}\right)^{m^2-r-1}+(-1)^{n+j}\left(-\frac{t}{n}\right)^{2m^2+r}\right]e^{-t}.
\end{align}

Then the integral looks similar to the Hankel representation of the gamma function:
\begin{align}
    \Gamma(z)=-\left[\oint_{\mathcal H}\frac {\dd t}{2\pi i} (-t)^{-z}e^{-t}\right]^{-1}.
\end{align}
Using this equivalence, the integral becomes
\begin{align}
    B_{n,j}^1\sim\frac{c_{n,j,D,m^2}^1}{(r+1)_{(n)}}\frac{n^{j+\frac{D}{2}-3-2m^2-r}}{(j+\frac{D}{2}-2)!\log(n)^{\frac{D-2}{2}}}\left[\frac{n^{m^2+2r+1}}{\Gamma(-m^2+r+1)}+\frac{(-1)^{n+j}}{\Gamma(-2m^2-r)}\right],\label{eq:gamma}
\end{align}
and substituting in (\ref{eq:normalization}) for $c_{n,j,D,m^2}^1$ yields the desired formula. When $2r+m^2+1> 0$, the first term in the brackets dominates the second as $n$ becomes large. The first term is positive when $r> m^2-1$, which gives positivity in this limit for a very large region of parameter space. 

\hspace{0pt}

\begin{figure}[ht]
    \centering
    \includegraphics[width=12cm]{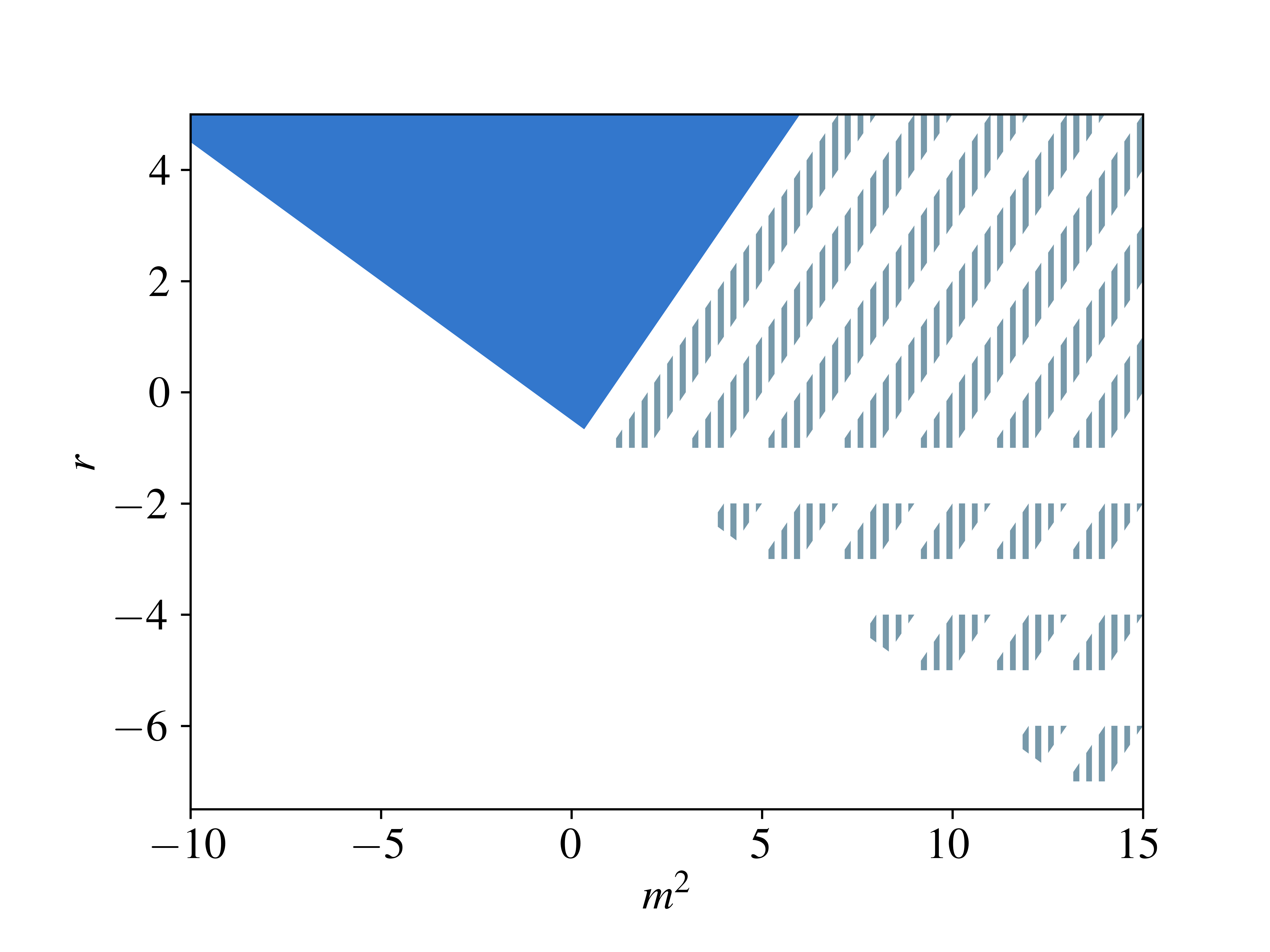}
    \caption{White denotes regions of $(r,m^2)$ space that violate unitarity in all $D$ by theorems \ref{theorem1}, \ref{theorem2}, and \ref{theorem9}. The slim gray regions denote areas we have not ruled out rigorously, but are nonetheless expected to violate unitarity in all $D$ by conjecture \ref{conjec}, and they can also be ruled out numerically. Blue denotes a region that is not ruled out analytically, but portions of it can be ruled out numerically in particular $D$ as shown in Fig. \ref{red}.}
    \label{fig:negative}
\end{figure}

The expression (\ref{eq:gamma}) also becomes negative when $(r+1)_{(n)}\geq 0$, $2r+m^2+1> 0$, and $k< m^2-r<k+1$ for some odd integer $k$. Additionally, when $m^2-r\in\mathbb N$, the sign can be positive or negative depending on the other parameters. This provides a new manifestly-negative region that differs from the one carved out by theorems \ref{theorem1} and \ref{theorem2}. The superposition of all negative regions from theorems \ref{theorem1}, \ref{theorem2}, and \ref{theorem9} is depicted in Fig. \ref{fig:negative}. 

\subsection{Regge Asymptotics}

In the Veneziano limit $r=0$, \cite{Arkani-Hamed:2022gsa} uses the double contour formula to demonstrate positivity of $B_{n,j}^1$ in the limit $n\to\infty$ with $n-j$ held fixed. Here we will generalize this proof to arbitrary $r$ and $m^2$. When $n-j$ is even, our proof is a straightforward extension of the one presented in \cite{Arkani-Hamed:2022gsa}; for completeness, we will still present this proof in its entirety. When $n-j$ is odd, the analysis becomes more complicated, as unlike the Veneziano case, the hypergeometric coefficients at odd $n-j$ are generally nonzero.

\hspace{0pt}

\begin{theorem}
    As $n\to\infty$ with $\Delta=n-j$ held fixed, we have the asymptotic relation for even $\Delta$:
   \begin{align}
    B_{n,n-\Delta}^1\sim \Gamma(r+1)\frac{\Gamma(\frac{D-3}{2})2^{-2n-\frac{1}{2}}e^{n-3m^{2}}}{(\frac{\Delta}{2})!\sqrt{\pi}n^{\frac{D}{2}-2+r}}\left(\frac{n}{3}\right)^{\Delta/2}
\end{align}
and for odd $\Delta$:
\begin{align}
    B_{n,n-\Delta}^1\sim(2r+m^2+1) \Gamma(r+1)\frac{\Gamma(\frac{D-3}{2})2^{-2n-\frac{3}{2}}e^{n-3m^{2}}}{(\frac{\Delta-1}{2})!\sqrt{\pi}n^{\frac{D}{2}-2+r}}\left(\frac{n}{3}\right)^{(\Delta-1)/2}
\end{align}
\end{theorem}

\textit{Proof.} To begin, we can relabel the integration variables $u$ and $v$ in the contour formula in order to re-express it as a commutator of two contour integrals:
\begin{align}
    B^1_{n,j}=\frac{c^1_{n,j,D,m^2}}{(r+1)_{(n)}}\left[\oint_{u=0}\frac{\dd u}{2\pi i}\oint_{v=0}\frac{\dd v}{2\pi i}-\oint_{v=0}\frac{\dd v}{2\pi i}\oint_{u=0}\frac{\dd u}{2\pi i}\right]\frac{(v-u)^j e^{v(2m^2+r+1)+u(m^2-r)}}{(uv)^{J+1}(e^{v}-e^{u})^{n+1}}.
\end{align} 
Then using the following equivalence familiar from conformal field theory,
\begin{align}
    \oint_{v=0}\oint_{u=0}-\oint_{u=0}\oint_{v=0}=\oint_{u=0}\oint_{v=u},\label{contourcommutator}
\end{align}
the expression for $B^1_{n,j}$ simplifies to\footnote{As noted in \cite{Arkani-Hamed:2022gsa}, this formula and the results we derive from it are also valid at odd $D$.}
\begin{align}
    B^1_{n,j}=\frac{-c^1_{n,j,D,m^2}}{(r+1)_{(n)}}\oint_{u=0}\frac{\dd u}{2\pi i}\oint_{v=u}\frac{\dd v}{2\pi i}\frac{(v-u)^j e^{v(2m^2+r+1)+u(m^2-r)}}{(uv)^{J+1}(e^{v}-e^{u})^{n+1}}.
\end{align}
Define $\Delta=n-j$ and rewrite:
\begin{align}
    B^1_{n,n-\Delta}=\frac{-c^1_{n,j,D,m^2}}{(r+1)_{(n)}}\oint_{u=0}\frac{\dd u}{2\pi i}\oint_{v=u}\frac{\dd v}{2\pi i}\left(\frac{u^{-1}-v^{-1}}{e^{v}-e^{u}}\right)^{n+1} \frac{e^{v(2m^2+r+1)+u(m^2-r)}}{(uv)^{\frac{D-2}{2}}(u^{-1}-v^{-1})^{\Delta+1}}.
\end{align}
We Taylor expand around the point $v=u$:
\begin{align}
    \frac{u^{-1}-v^{-1}}{e^v-e^u}=\frac{e^{-u}}{u^2}\left(1-\frac{u+2}{2u}(v-u)+\frac{u^2+6u+12}{12u^2}(v-u)^2+\mc O((v-u)^3)\right).
\end{align}
Now as we send $n\to\infty$, one can write
\begin{align}
    \left(\frac{u^{-1}-v^{-1}}{e^{v}-e^{u}}\right)^{n+1}\sim  \frac{e^{-u(n+1)}}{u^{2n+2}}\exp[\frac{n+1}{u^2}\left(-\frac{u+2}{2u}(v-u)+\frac{u^2+6u+12}{12u^2}(v-u)^2+\mc O((v-u)^3)\right)]
\end{align}
and see that we obtain a rapidly oscillating exponential as $u$ circles around $0$. As pointed out in \cite{Arkani-Hamed:2022gsa}, one can verify that the terms of order $(v-u)^3$ and higher will not contribute to the final result, so we will ignore them from here. Next we can choose the $u$ contour to be a circle of radius $2$ around $u=0$, and then evaluate it via a stationary-phase approximation around the saddle point $u=-2$. Substituting $u=-2$ and taking only the lowest power of $(v-u)^{-1}$ outside the exponential, we get
\begin{align}\begin{split}
    B^1_{n,n-\Delta}\sim\frac{-c_{n,j,D,m^2}^1}{(r+1)_{(n)}}\oint_{u=0}&\frac{\dd u}{2\pi i}\oint_{v=u}\frac{\dd v}{2\pi i}\frac{e^{-6m^2-2-u(n+1)}}{u^{2n-2\Delta+D-2}(v-u)^{\Delta+1}}\\ &\times \exp[(n+1)\left(\frac{u+2}{4}(v-u)+\frac{1}{12}(v-u)^2\right)].
    \end{split}
\end{align}
Notably, in this step all dependence on $r$ and $m^2$ reduces down to a simple multiplicative factor. We make a change of variables $t=(v-u)\sqrt{n+1}$ and reorder the integrals:
\begin{align}\begin{split}
    B_{n,n-\Delta}^1\sim\frac{-c_{n,j,D,m^2}^1 (n+1)^{\frac{\Delta}{2}}}{(r+1)_{(n)}}\oint_{t=0}&\frac{\dd t}{2\pi i}\oint_{u=0}\frac{\dd u}{2\pi i}\frac{e^{-6m^2-2-u(n+1)}}{u^{2n-2\Delta+D-2}t^{\Delta+1}}\\&\times\exp[\frac{(u+2)\sqrt {n+1}}{4}t+\frac{1}{12}t^2].\end{split}
\end{align}
Evaluate the $u$ integral, then approximate $2n-2\Delta+D-2\sim 2n$ in the integrand to get
\begin{align}
    B_{n,n-\Delta}^1\sim \frac{c_{n,j,D,m^2}^1 (n+1)^{2n-\frac32\Delta+D-3}e^{-6m^2-2}}{(2n-2\Delta+D-3)!(r+1)_{(n)}}\oint_{t=0}\frac{\dd t}{2\pi i}\exp[\frac{\sqrt{n+1}}{2}t+\frac{t^2}{12}] \frac{(1-\frac{t}{4\sqrt {n+1}})^{2n}}{t^{\Delta+1}}.\label{eq:tintegral}
\end{align}
Next we use the relation 
\begin{align}
    \left(1-\frac{t}{4\sqrt {n+1}}\right)^{2n}=e^{-\frac{t\sqrt {n+1}}{2}-\frac{t^2}{16}}+\mc O(n^{-1/2}),
\end{align}
which can be seen by writing $(1-\alpha)^n=\exp[n\log (1-\alpha)]$ and expanding the logarithm as a Taylor series. Substituting this into (\ref{eq:tintegral}) gives
\begin{align}
    B_{n,n-\Delta}^1\sim  \frac{c_{n,j,D,m^2}^1 (n+1)^{2n-\frac32\Delta+D-3}e^{-6m^2-2}}{(2n-2\Delta+D-3)!(r+1)_{(n)}}\oint_{t=0}\frac{\dd t}{2\pi i} \frac{ e^{\frac{t^2}{48}}}{t^{\Delta+1}}.\label{eq:oint}
\end{align}
When $\Delta$ is even, evaluating the integral yields
\begin{align}
    B_{n,n-\Delta}^1\sim \frac{c_{n,j,D,m^2}^1 (n+1)^{2n-\frac32\Delta+D-3}e^{-6m^2-2}}{48^{\Delta/2}(\frac{\Delta}{2})!(2n-2\Delta+D-3)!(r+1)_{(n)}}.
\end{align}
Simplifying this via Stirling's approximation gives
\begin{align}
    B_{n,n-\Delta}^1\sim \frac{c_{n,j,D,m^2}^1}{(r+1)_{(n)}} \frac{2^{-2n-D+2}e^{2n-6m^2}}{(\frac{\Delta}{2})!\sqrt{\pi n}}\left(\frac n3\right)^{\Delta/2},
\end{align}
and the final result can be obtained by substituting in the value of $c_{n,j,D,m^2}^1$ and applying Stirling's approximation a second time. On the other hand, when $\Delta$ is odd, (\ref{eq:oint}) simply evaluates to zero, indicating that the odd coefficients are smaller than the even ones by at least a factor of $\sqrt n$. To find a more precise expression for these coefficients, we follow a similar procedure to the one used in Section \ref{sec:even}. When $n-j$ is odd, we can let $2r+m^2+1=pq$ for some $p\in\mathbb N$ and $q\in [-1,1]$, and then factor the integrand of the contour formula as
\begin{align}
    B^1_{n,j}=\frac{c^1_{n,j,D,m^2}}{(r+1)_{(n)}}\oint_{u=0}\frac{\dd u}{2\pi i}\oint_{v=0}\frac{\dd v}{2\pi i}(e^{qv}-e^{qu})\sum_{k=0}^{p-1}\frac{(v-u)^j e^{(qk+m^2-r)v+(q(p-1-k)+m^2-r)u}}{(uv)^{J+1}(e^{v}-e^{u})^{n+1}}.
\end{align}
Expanding the factor of $e^{qv}-e^{qu}$ in powers of $t=(v-u)\sqrt n$, the dominant contribution to the integral is given by the highest order term $qe^{qu}(v-u)$. Substituting this in, we find the expression 
\begin{align}
    B^1_{n,n-\Delta}\sim q\sum_{k=0}^{p-1}\frac{c^1_{n,j,D,m^2}}{(r+1)_{(n)}}\oint_{u=0}\frac{\dd u}{2\pi i}\oint_{v=0}\frac{\dd v}{2\pi i}\frac{(v-u)^{j+1} e^{(qk+m^2-r)v+(q(p-k)+m^2-r)u}}{(uv)^{J+1}(e^{v}-e^{u})^{n+1}},
\end{align}
which is just $q$ times a sum over the even $\Delta$ coefficients, evaluated at different values of $r$. As $n\to\infty$ with $\Delta$ held fixed, we have already shown that this integral has no dependence on $r$, so all terms in the sum will asymptote to the same value. Therefore we can replace the sum with a simple prefactor of $pq/2=(2r+m^2+1)/2$, yielding
\begin{align}
    B^1_{n,n-\Delta}\sim\frac{2r+m^2+1}2 B_{n,n-\Delta+1},\label{eq:regge_odd}
\end{align}
where we have used that $c^1_{n,n-\Delta,D,m^2}/c^1_{n,n-\Delta+1,D,m^2}\sim 1$ as $n\to\infty$. This is equivalent to the desired result. We also see that the sign of the coefficients for any $\Delta$ is characterized by
\begin{align}
    B_{n,n-\Delta}^1\propto (2r+m^2+1)^\Delta\Gamma(r+1).
\end{align}

\section{Conclusion}
\label{sec:conclusion}

We have established a subset of the hypergeometric amplitude's parameter space where positivity is satisfied, and another subset where positivity is violated. By directly examining the formula for the residues's partial wave coefficients as an inner product against a Gegenbauer polynomial, we were able to carve out an infinite region where positivity is violated in all dimensions, and establish that in the remaining region, any given coefficient becomes positive when $r$ is sufficiently large. Then by developing a contour integral representation of the coefficients, we were able to demonstrate that all coefficients are positive in a different infinite region of parameter space. This formula also allowed for computation of various asymptotic expressions for the coefficients, which in turn demonstrated a new region where positivity is violated. A summary of all currently known parameter bounds on positive coefficients is presented in Fig. \ref{summaryfig}.

\begin{figure}
\begin{tcolorbox}[
    title= \textbf{Summary of Positivity Bounds}]

    \hspace{0pt}

    There exists an $n$, $j$ such that $B^1_{n,j}< 0$ if any of the following hold (see Fig. \ref{fig:negative}):
    \begin{enumerate}
    \itemsep0em 
        \item $\lfloor r\rfloor$ is negative and even.
        \item $r<-\frac{1+m^2}{2}$.
        \item $m^2-r>1$ and $\lceil 6m^2\rceil$ is odd.
        \item $k<m^2-r<k+1$ for some odd $k\in\mathbb N$.
    \end{enumerate}
    When none of these conditions hold, we can guarantee that $B^1_{n,j}\geq 0$ whenever any of the following are true:
    \begin{enumerate}
\itemsep0em 
        \setcounter{enumi}{4}

    \item $n \leq 2+2r+m^2+ 2j(2H_{2n-3}-H_{n-1})^{-1}$ where $H_k$ is the $k$th harmonic number (see Fig. \ref{redblue}).

    \item $m^2\leq \frac{4-\tilde D}{6}$ and $n-j+1\geq 2^k\geq  2m^2+r+\frac{\tilde D-2}{4}$ for some $k\in\mathbb N$ (see Fig. \ref{fig:green}). Here $\tilde D=2\lfloor \frac D2\rfloor$. The restriction on $n-j$ is trivially satisfied in a large region of $(r,m^2,D)$ space, giving positivity of the full amplitude.
    
    \item $m^2\leq \frac{6-\tilde D}6$, $n-j$ is odd, and $n-j\geq 2^k\geq 2m^2+r+\frac{\tilde D-2}{4}$ for some $k\in\mathbb N$ (see Fig. \ref{fig:odd})

    \item $n-j$ is even, $D=4$, and $2m^2+r\leq -\frac{1}{\sqrt 6}$ (see Fig. \ref{orangepurple}).

    \item $n-j$ is even, $D\geq 6$, $ 2m^2+r+\frac{\tilde D}{4}\leq 1$, and \begin{align*}
        \frac{j+1}{n+1}&\leq \frac{G_{\tilde D-4}}{G_{\tilde D-5} \left(2m^2+r+\frac{\tilde D-2}{4}\right)}-1
    \end{align*}
    where $G_m$ is the absolute value of the $m$th Gregory coefficient (see Fig. \ref{orangepurple}). 

    \item $j=0$, $D=4$, and $r\geq m^2$. There is also a general procedure for establishing a positive region at any particular $j$ and $D$ with a finite amount of work.

    \item $D\leq 26$ and $m^2=-1$, or $D\leq 10$ and $r=m^2=-\frac13$, by appealing to the no-ghost theorem.
\end{enumerate}

Lastly, we established that as we send $n\to \infty$, $B^1_{n,j}$ and $B^1_{n,n-\Delta}$ both remain non-negative when $r\geq\max(-\frac{m^2+1}{2},m^2+1)$.

\end{tcolorbox}
\caption{Summary of Positivity Bounds.}\label{summaryfig}
\end{figure}

\hspace{0pt}

Various special cases of items 1 and 2 have been demonstrated via a mixture of numerical and analytical methods in \cite{Rigatos:2023asb, Rigatos:2024beq, Wang:2024wcc, Cheung:2024uhn}, but the complete analytical derivations at general $r,m^2$ that we have presented here, as well as items 3 and 4, are new. Most of the ``positive'' results are also new, with a few exceptions. Positivity in $D\leq 26$ at $m^2=-1$ is straightforward and has been discussed in \cite{Rigatos:2023asb}. A restricted version of items 6 and 10 was proved in \cite{Arkani-Hamed:2022gsa} for the two special cases where $m^2=-1,r=0$ and $m^2=-\frac13,r=-\frac13$, which correspond to the bosonic and type-I Veneziano amplitude respectively. The paper \cite{Arkani-Hamed:2022gsa} also proved positivity in the large $n$, fixed $j$ and large $n$, fixed $n-j$ limits for these two special cases. The large-$n$, fixed-$j$ asymptotics at $r=0$ for arbitrary $m^2$ are also calculated in \cite{Bhardwaj:2024klc}. Lastly, a generalized version of item 10 was demonstrated in \cite{Rigatos:2024beq, Wang:2024wcc}. Apart from these, all other listed results have not yet been proven elsewhere.

\hspace{0pt}

There is much opportunity for further work on the hypergeometric amplitude. While we have sculpted out a sizeable positive region of parameter space, the region we have found is still quite far off from the numerical results depicted in Fig. \ref{red}.\footnote{In fact, our analytic bounds almost never quite meet the numerical predictions for the critical dimension, save for a small region in $D=4$ given by theorem \ref{theorem6}, located near $(r,m^2)\approx (-1/2,0)$.} A stronger version of lemma 2 would be one immediate avenue for enlargening the positive region, and could also prove the positivity of the bosonic and type-I Veneziano amplitudes in $D=12$ and $D=8$ respectively. Alternatively, a proof that the positive region is exclusively determined exclusively by the low-spin coefficients $B_{n,j}^1$ as conjectured in \cite{Rigatos:2024beq, Wang:2024wcc} could dramatically enlarge the positive region, perhaps entirely eclipsing the results presented here. Additionally, as depicted in Fig. \ref{fig:negative}, there is an infinite family of potentially unitary ``islands" at positive $m^2$ that we were unable to rule out rigorously; future work could involve locating a negative coefficient in each of these areas. This could potentially be done by proving the conjecture \ref{conjec} or by examining large-$n$ asymptotics where $j$ varies as some carefully-chosen function of $n$. Yet another avenue for further research would be a generalization of these results to the $q$-hypergeometric amplitude introduced in \cite{Cheung:2023adk}, whose positivity has also been explored by \cite{Wang:2024wcc}. In \cite{Bhardwaj:2022lbz}, the authors perform work on a contour representation for the Coon amplitude similar to the one we presented here. It could be interesting to search for a contour representation that holds when both $q$ and $r$ are nonzero. Lastly, more amplitudes that exhibit integer spectra have been explored in \cite{Haring:2023zwu, Cheung:2024uhn}; it would be interesting to see if a similar contour representation can demonstrate positivity for these amplitudes.

\hspace{0pt}

The hypergeometric amplitude is also a source of interesting problems beyond the question of positivity. The complete consistency of the hypergeometric amplitude demands a generalization to $n$-particle scattering that correctly factorizes onto lower point amplitudes; this constraint has recently ruled out several other proposed UV-complete amplitudes \cite{Arkani-Hamed:2023jwn}, and explicit factorization constraints for integer-spectrum amplitudes such as the hypergeometric have been worked out in \cite{Gross:1969db}. Construction of such a generalization would be the next natural step in demonstrating complete consistency of the hypergeometric amplitude. In addition, \cite{Cheung:2023adk} also constructs a hypergeometric generalization of the Virasoro-Shapiro amplitude. The Virasoro-Shapiro and Veneziano amplitudes satisfy the KLT relation
\begin{align}
    A_{\rm vs}=\frac{\sin(\pi s)\sin(\pi t)}{\pi\sin(\pi(-s-t))}A_{\rm ven}^2.
\end{align}
An interesting question raised in \cite{Rigatos:2023asb} would be to see if an analogous relation holds between the hypergeometric variant of each amplitude.

\section*{Acknowledgements}
We are grateful to Rishabh Bhardwaj, Shounak De, Grant Remmen, and Anastasia Volovich for many helpful discussions. This work was supported in part by the US Department of Energy under contract DE-SC0010010 Task F.  MS is also grateful for support by the Munich Institute for Astro-, Particle and BioPhysics (MIAPbP) which is funded by the Deutsche Forschungsgemeinschaft (DFG, German Research Foundation) under Germany's Excellence Strategy -- EXC-2094 -- 390783311, and thanks the Galileo Galilei Institute for Theoretical Physics for the hospitality and the INFN for partial support during the completion of this work.

\appendix

\section{Regge Coefficients}
\label{sec:regge}

We can expand the polynomial in (\ref{eq:gegen}) as follows:
\begin{align}\begin{split}
    &\pdv[j]{x} \left[\frac{n-3m^2}{2}x+\frac{m^2-n}{2}+r+1\right]_{(n-1+a)}\\
    &=\sum_{k=0}^{\Delta} \left(\frac{n-3m^2}{2}\right)^{k+n-\Delta-1+a} x^{k} (k+1)_{(n-\Delta-1+a)}\sum_{1\leq \ell_1<\ldots<\ell_{\Delta-k}}^{n-1+a} \prod_{i=1}^{\Delta-k} \left(\frac{m^2-n}{2}+r+\ell_i\right)\end{split}
\end{align}
where $\Delta=n-1+a-j$. This formula can be used to explicitly integrate (\ref{eq:gegen}) at fixed $\Delta$, yielding general expressions for coefficients along Regge trajectories. Defining $w=2r+m^2+a$, we compute

\begingroup
\allowdisplaybreaks
\begin{align*}
    B^a_{n,n-1+a}&\propto\frac{(n-3m^2)^{n-1+a}}{(r+1)_{(n)}}\\
    B^a_{n,n-2+a}&\propto \frac{(n-3m^2)^{n-2+a}}{(r+1)_{(n)}}w\\
    B^a_{n,n-3+a}&\propto\frac{(n-3m^2)^{n-3+a}}{(r+1)_{(n)}}\left[3 w^2 - (n+a) + \frac{3 (n-3m^2)^2}{2(n+a) + D-7}\right]\\
    B^a_{n,n-4+a}&\propto\frac{(n-3m^2)^{n-4+a}}{(r+1)_{(n)}}w\left[ w^2  - (n+a)  + 3 \frac{(n-3m^2)^2}{ 2(n+a) + D-9}\right]\\
    B^a_{n,n-5+a}&\propto \frac{(n-3m^2)^{n-5+a}}{(r+1)_{(n)}}\Bigg[15 w^4 + (n+a)(2- 30 w^2 + 5 (n+a)) \\&+ 
 30 \frac{( n-3m^2)^2 (3 w^2 - (n+a))}{2(n+a) + D-11} + \frac{
 45 (n-3m^2)^4}{( 2 (n+a) + D-11) (2 (n+a) + D-9)}\Bigg]\\
     B^a_{n,n-6+a}&\propto \frac{(n-3m^2)^{n-6+a}}{(r+1)_{(n)}}w\Bigg[3 w^4 + (n+a) (2 -10w^2+ 5 (n+a))
\\& + 
 \frac{30  (n-3m^2)^2(w^2 - (n+a))}{2(n+a)+D-13} + \frac{45 ( n-3m^2)^4}{(2 (n+a) + D-13)(2(n+a)+D-11)}\Bigg]  \tag{\stepcounter{equation}\theequation } 
\end{align*}
\endgroup

\section{Lemmas for Even Coefficients}
\label{sec:appendix}

\subsection{Proof of Lemma 1}

\textbf{Lemma 1.} \textit{The $a_k$ coefficients are defined by} 
    \begin{align}
        \frac{(1-z)^p}{(-\log(1-z))^q}=\sum_{k=-3}^\infty a_k(p,q)z^k
    \end{align}
    \textit{When $k\geq-q+2$ and one of the following is true, }
    \begin{enumerate}
        \item $q=1$ and $p\leq -\frac1{\sqrt 6}$
        \item $q=2$ and $p\leq \frac{-3 - \sqrt 3}{6}$
        \item $q\geq 3$ and $ p\leq \frac{1-q}2$
    \end{enumerate}
    \textit{then $a_k\geq 0$ and $2a_{k+1}\geq a_k$.}

\hspace{0pt}

\textit{Proof.} We will prove by induction on $q$. First, positivity in the base case $q\in\{1,2,3\}$ can be proven by a method used in \cite{Arkani-Hamed:2022gsa}; to save space we will only briefly summarize this argument in one case: when $q=3$, the coefficients admit an integral representation,
\begin{align}\begin{split}
    a_{k-3}(p,3)=\frac1{2k!}\int_0^1\dd s\> (-p - s)_{(k-2)}\Big[&k^2 s^2 - k(s^2+ (1 + p + s) (1 + 
      2 s))  \\&
   +2 (1 + p + s) (2 + p + s)  \Big].\end{split}\label{eq:inta}
\end{align}
From here one can use the fact that the positive term of order $k^2$ dominates over the others when $k$ is sufficiently large to prove that $0 \leq a_{k}(p,3)$ and $2a_{k+1}(p,3)-a_k(p,3)\geq 0$ for sufficiently large $k$, and then manually verify that it holds for smaller $k$. Some worked examples of this argument, with values of $q$ and $p$ that differ from ours, are present in \cite{Arkani-Hamed:2022gsa}. 

\hspace{0pt}

Now we proceed by induction on $q$. Assume that $2a_{k+1}\geq a_k(p,q)\geq 0$ for all $p\leq (1-q)/2$ and $k\geq -q+2$ at some $q$. We have the identity
\begin{align}
    \frac{z(1 - z)^{p}}{(-\log(1 - z))^{q+1}}=\int_0^1\dd s\> \frac{(1 - z)^{p + s}}{(-\log(1 - z))^q},
\end{align}
which implies the following relation among the $a_k(p,q)$ coefficients:
\begin{align}
   a_k\left(p,q+1\right)= \int_0^1\dd s\> a_{k+1}\left(p+s,q\right).\label{eq:inta2}
\end{align}
By the inductive hypothesis, the integrand will be positive for all $s< \frac12$ if $p\leq q-\frac12$. The integrand is a polynomial in $s$, and since all roots lie at $s\geq \frac12$, the  integral must be positive. Therefore $a_k(p,q+1)$ is positive as long as $k\geq -(q+1)+2$ and $p\leq \frac{1-(q+1)}{2}$, which is what we wanted. A similar argument shows that $2a_{k+1}(p,q+1)- a_k(p,q+1)\geq 0$ for $p\leq \frac{1-(q+1)}{2}$.

\subsection{Proof of Lemma 2}

\label{sec:lemma2}

\textbf{Lemma 2.}\textit{
   Consider the Laurent expansion first about $y=0$, then about $x=0$:}
   \begin{align*}
        \frac{\left(\frac{1}{\log(1-x)}-\frac1{\log(1-y)}\right)^j}{(x-y)^n}&=\sum_{\alpha,\beta}b_{\alpha,\beta}(n,j)x^\alpha y^\beta
   \end{align*}
   \textit{When $\alpha+\beta+j+n\neq 0$, the $b_{\alpha\beta}$ coefficients satisfy}
    \begin{align*}
    \frac{n-1+\beta}{j+\beta}\frac{G_{\lceil(\alpha+n)/j\rceil+2}}{G_{\lceil(\alpha+n)/j\rceil+1}}b_{\alpha,\beta+1}\geq \frac{G_{\lceil(\alpha+n)/j\rceil+2}}{G_{\lceil(\alpha+n)/j\rceil+1}}b_{\alpha+1,\beta}\geq b_{\alpha,\beta}\geq 0
\end{align*}

\textit{Proof}. We begin by computing an explicit formula for $b_{\alpha\beta}(n,j)$ as a combinatorial sum. Substitute in the Laurent series for $1/\log(1-z)$:
\begin{align}\begin{split}
    \left(\frac{\frac1{\log(1-x)}-\frac1{\log(1-y)}}{x-y}\right)^j&=\left(-\frac{x^{-1}-y^{-1}}{x-y}+\sum_{m=0}^\infty G_{m+1} \frac{x^m-y^m}{x-y}\right)^j\\&=\left(\frac1{xy}+\sum_{m=1}^\infty\sum_{k=0}^{m-1} G_{m+1} x^{m-1-k}y^k\right)^j.\end{split}\label{parj}
\end{align}
Every term here is positive, giving the first part of the lemma that $b_{\alpha\beta}\geq 0$. Now to evaluate the power of $j$, we use the multinomial theorem and find that the coefficient on $x^\gamma y^\delta$ is given by
\begin{align}
c_{\gamma,\delta}=\sum_{(m)^{\gamma+j},(k)^{\delta+j}}\prod_{i=1}^j  G_{m_i+k_i}\label{eq:mz}
\end{align}
where the inner sum runs over all pairs $(m,k)$ of ordered length-$j$ sequences of nonnegative integers satisfying $\sum_i m_i=\gamma+j$ and $\sum_\ell k_\ell=\delta+j$ (these are known as the \textit{weak compositions into $j$ parts} of $\gamma+j$ and $\delta+j$). Now for a positive integer $z$, let $M_{z}$ denote the set of weak compositions of $z$ into $j$ parts, so that (\ref{eq:mz}) is a sum over the elements of $M_{\gamma+j}\times M_{\delta+j}$. We can define an injective map $f:M_{z}\to M_{z+1}$ which, for a given sequence $m^z\in M_{z}$, picks out a maximum element $m^z_{\rm max}\in m^z$ and increments it by 1, and keeps all other elements the same. This gives the bound
\begin{align}
    \sum_{(m)^z,(k)}\prod_{i} G_{m^z_i+k_i} \geq \sum_{(m)^{z-1},(k)}\prod_{i}  G_{f(m^{z-1})_i+k_i} =\sum_{(m)^{z-1},(k)}\frac{ G_{m^{z-1}_{\rm max}+1+k_{\rm max}}}{ G_{m^{z-1}_{\rm max}+k_{\rm max}}}\prod_{i}  G_{m^{z-1}_i+k_i}
\end{align}
where the first inequality follows from injectivity of $f$. 
Next, we observe that $m^{\gamma+j}_{\rm max}\geq \lceil 1+\frac{\gamma}{j}\rceil$, since $m^z$ must contain at least one term greater than or equal to its average. Plugging this back in gives the inequality
\begin{align}
c_{\gamma+1,\delta}\geq\frac{G_{\lceil\gamma/j \rceil+3}}{G_{\lceil\gamma/j\rceil+2}}c_{\gamma,\delta}.\label{ccoef}
\end{align}
Now multiplying (\ref{parj}) by $(x-y)^{j-n}$, we get the expression for $b_{\alpha\beta}$
\begin{align}
    b_{\alpha\beta}(n,j)=\sum_{\ell=0}^{\beta+j} c_{\alpha +\beta+n-\ell,\ell-j} \binom{n-1+\beta-\ell}{j+\beta-\ell}.
\end{align}
Then plugging in (\ref{ccoef}) and applying some inequalities between the binomial coefficients obtains the desired result so long as $\alpha+\beta+j+n\neq 0$:
\begin{align}
    b_{\alpha+1,\beta}\geq \frac{G_{\lceil(\alpha+n)/j\rceil+2}}{G_{\lceil(\alpha+n)/j\rceil+1}}b_{\alpha,\beta}\qq{and}
    b_{\alpha,\beta+1}\geq \frac{n-1+\beta}{j+\beta}b_{\alpha+1,\beta}.
\end{align}

\bibliographystyle{JHEP} 
\bibliography{JHEP}

\end{document}